\documentclass[
final
, nomarks
]{dmtcs-episciences}

\usepackage[utf8]{inputenc}
\usepackage{latexsym,algorithm2e,enumerate}
\usepackage{amsmath,amssymb,amsthm,mathrsfs,amsfonts}
\usepackage{tabularx}
\usepackage{multirow}
\usepackage{booktabs}

\newtheorem{theorem}{Theorem}
\newtheorem{corollary}[theorem]{Corollary}
\newtheorem{lemma}[theorem]{Lemma}
\newtheorem{proposition}[theorem]{Proposition}

\def\D{\mathscr{D}}

\def\C{\mathscr{C}}

\usepackage[round]{natbib}

\author{F\'abio~Protti\affiliationmark{1}
  \and U\'everton S. Souza\affiliationmark{1}}
\title[Decycling a graph by the removal of a matching]{Decycling a graph by the removal of a matching: new algorithmic and structural aspects in some classes of graphs}
\affiliation{
Instituto de Computa\c c\~ao, Universidade Federal Fluminense, Niter\'oi, RJ, Brazil}
\received{2017-10-19}

\revised{2018-8-21}

\accepted{2018-10-24}

\begin{document}
\publicationdetails{20}{2018}{2}{15}{3998}
\maketitle
\begin{abstract}
A graph $G$ is {\em matching-decyclable} if it has a matching $M$ such that $G-M$ is acyclic. Deciding whether $G$ is matching-decyclable is an NP-complete problem even if $G$ is 2-connected, planar, and subcubic. In this work we present results on matching-decyclability in the following classes: Hamiltonian subcubic graphs, chordal graphs, and distance-hereditary graphs. In Hamiltonian subcubic graphs we show that deciding matching-decyclability is NP-complete even if there are exactly two vertices of degree two. For chordal and distance-hereditary graphs, we present characterizations of matching-decyclability that lead to $O(n)$-time recognition algorithms.
\end{abstract}

\section{Introduction}

In this work we focus on the following problem: given a graph $G$, is it possible to destroy all of its cycles by removing a matching from its edge set? Equivalently, is it possible to find a partition $(M,F)$ of $E(G)$ such that $M$ is a matching and $F$ is acyclic? If the answer is ``yes'' then we say that $M$ is a {\em decycling matching} of $G$, and $G$ is a {\em matching-decyclable graph}, or simply {\em m-decyclable}.

The problem of destroying all the cycles of a graph by removing a set of edges (a {\em decycling set}) has already been considered. For a graph $G$ on $n$ vertices and $m$ edges and with $w$ connected components, a minimum decycling set $E^*$ has exactly $m-n+w$ edges, because the removal of $E^*$ must leave a spanning forest of $G$. On the other hand, for directed graphs, finding a minimum set of arcs whose removal leaves an acyclic digraph is precisely the optimization version of the classical Feedback Arc Set Problem, a member of Karp's list of $21$ NP-complete problems~\citep{karp72}.

M-decyclable graphs have recently been studied in~\citep{lima16}, where the authors prove that recognizing match\-ing-decyclability is NP-complete even for $2$-connected planar fairly cubic graphs. (A graph is {\em fairly cubic} if it has $n-2$ vertices of degree three and two vertices of degree two.) The authors also show polynomial-time recognition algorithms of m-decyclable graphs restricted to chordal, $P_5$-free, (claw,paw)-free, and $C_4$-free distance hereditary graphs, but no structural characterizations of m-decyclable graphs in such classes are provided.

As we shall see later in this work, a necessary (but not sufficient) condition for a graph $G$ to be m-decyclable is that $|E(H)|\leq\lfloor\,\frac{3}{2}|V(H)|\,\rfloor-1$ for every subgraph $H$ of $G$. Say that a graph $G$ satisfying such a necessary condition is {\em sparse}. Two-connected fairly cubic graphs are sparse, thus the NP-completeness result in~\citep{lima16} tells us that deciding matching-decyclability is hard even for a subset of sparse graphs. On the other hand, a natural question is to find graph classes in which being sparse is equivalent to being m-decyclable. In the next sections, we show that this is exactly the case for chordal graphs and $K_{2,4}$-free distance-hereditary graphs.

The remainder of this work is organized as follows. Section 2 contains the necessary background. In Section 3 we show that deciding whether a Hamiltonian fairly cubic graph is m-decyclable is NP-complete; this result strengths the result in~\citep{lima16}, since Hamiltonian fairly cubic graphs form a subclass of 2-connected fairly cubic graphs. In Section 4 we characterize m-decyclable chordal graphs; the characterization leads to a simple $O(n)$-time recognition algorithm for such graphs, refining a previous result presented in~\citep{lima16}. M-decyclable split graphs are also considered in Section 4. Section 5 describes a characterization of m-decyclable distance-hereditary graphs and a direct application of this result to cographs; the characterization extends the result in~\citep{lima16}, and implies a simple $O(n)$-time recognition algorithm. Section 6 contains our conclusions.

\section{Preliminaries}

In this work, all graphs are finite, simple, and nonempty. Let $G$ be a graph with $|V(G)|=n$ and $|E(G)|=m$. The degree of a vertex $v\in V(G)$ is denoted by $d_G(v)$. The minimum degree of $G$ is defined as $\delta(G)=\min\{d_G(v): v\in V(G)\}$. A {\em cut vertex} (resp., {\em bridge}) is a vertex (resp., edge) whose removal disconnects $G$. A {\em block} of $G$ is either a bridge or a maximal 2-connected subgraph of $G$. A {\em leaf block} is a block containing exactly one cut vertex. We say that $G$ {\em contains} $H$ if $H$ is a (not necessarily induced) subgraph of $G$. If, in addition, $H$ is induced, we say that $G$ {\em contains $H$ as an induced subgraph}. If $G$ does not contain $H_1, H_2,\ldots,H_k$ as induced subgraphs then $G$ is $(H_1,H_2,\ldots,H_k)${\em -free}.

\if 10

Two edges are {\em disjoint} if they have no endpoint in common. A matching in a graph $G$ is a set of mutually disjoint edges. Let $M$ denote a matching in a graph $G$. A vertex $v$ is {\em matched} if $v$ is an endpoint of some edge in $M$. If $uv\in M$ then $u$ and $v$ are {\em mates}. An {\em $M$-augmenting path} is a path that alternates edges in $M$ and $E(G)\setminus M$, and whose endvertices are not matched. It is well known~\citep{berge57} that if there is an $M$-augmenting path $P$ then $M^*=M\,\Delta\,E(P)$ is a matching satisfying $|M^*|=|M|+1$ (where the symbol $\Delta$ stands for the symmetric difference between two sets).

\fi

We say that $G$ is {\em subcubic} if all of its vertices have degree at most three, and {\em fairly cubic} if $G$ contains $n-2$ vertices of degree three and two vertices of degree two (the latter terminology is adopted from~\citep{chaea-et-al-07}, p. 2985). A graph $H$ is {\em bad} if $|E(H)|>\lfloor\,\frac{3}{2} |V(H)|\,\rfloor-1$. Say that $G$ is {\em sparse} if $G$ contains no bad subgraph. If $G$ is sparse then, of course, $m\leq\lfloor\,\frac{3}{2}n\,\rfloor-1$.

\begin{figure}[htbp]
\centering
\includegraphics[width=0.7\textwidth]{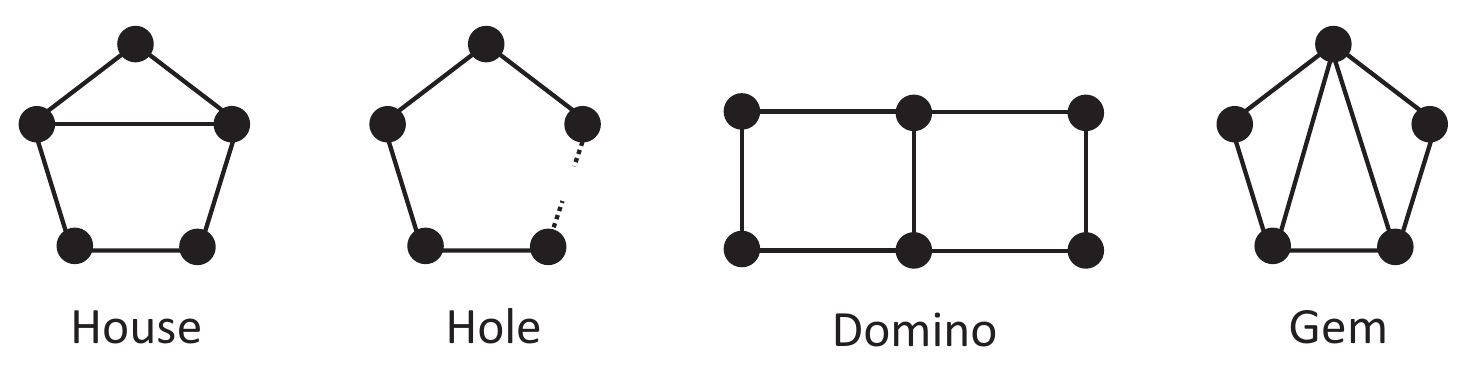}
\caption{Forbidden induced subgraphs for distance-hereditary graphs~\ref{lem:hc-edge}.}\label{fig:dhg}
\end{figure}

The complete graph with $n$ vertices is denoted by $K_n$. The graph $K_3$ is called {\em triangle}. A $2K_2$ is graph with vertices $a,b,c,d$ and edges $ab,cd$. A {\it gem} is a graph with vertices $a,b,c,d,e$ and edges $ab$, $bc$, $cd$, $ae$, $be$, $ce$, $de$. A {\it house} is a graph with vertices $a,b,c,d,e$ and edges $ab$, $bc$, $cd$, $ad$, $ae$, $be$. A {\it domino} is a graph with vertices $a,b,c,d,e,h$ and edges $ab$, $bc$, $cd$, $ad$, $be$, $eh$, $ch$. A {\em square} is a $4$-cycle with no chords. A {\em diamond} is a graph consisting of a $4$-cycle plus one chord. A $k$-{\em hole} (or simply {\em hole}) is a $k$-cycle with no chords, for $k\geq 5$.  We denote by $K_{3,3}^-$ the graph obtained by removing one edge of $K_{3,3}$, and by $P_k$ the path with $k$ vertices. A {\em chordal graph} is a (square, hole)-free graph. A {\em split graph} is a (square, $5$-hole, $2K_2$)-free graph~\citep{fh77}. A {\em cograph} is a $P_4$-free graph~\citep{corneil}. A {\em distance-hereditary graph} is a (house, hole, domino, gem)-free graph~\citep{bandelt86}. See Figure~\ref{fig:dhg}.

We say that $G$ is {\em m-decyclable} if there is a partition $(M,F)$ of $E(G)$ such that $M$ is a matching and $F$ is acyclic; in this case, $M$ is a {\em decycling matching} of $G$. It is easy to see that being m-decyclable is a property inherited by all subgraphs. This fact and other useful facts are listed in the proposition below; some of them are already mentioned in~\citep{lima16}.

\begin{proposition}\label{prop:super-prop} Let $G$ be a graph. Then:

$(a)$ If $G$ is m-decyclable then every subgraph of $G$ is m-decyclable.

$(b)$ If $G$ is m-decyclable then $m\leq\lfloor\,\frac{3}{2}n\,\rfloor-1$.

$(c)$ If $G$ is sparse then every subgraph of $G$ is sparse.

$(d)$ If $G$ is sparse then $G$ contains at least two vertices of degree two or less.

$(e)$ If $G$ is sparse then $G$ contains no $K_4$, $K_{3,3}$, or gem.

$(f)$ Every $2$-connected fairly cubic graph is sparse.

$(g)$ If $G$ is m-decyclable then $G$ is sparse.

$(h)$ The graph $K_{2,4}$ is not m-decyclable.

$(i)$ If $G$ is connected and matching-decyclable then $G$ has a matching $M$ for which $G-M$ is a tree.

$(j)$ If $G$ is subcubic and connected, then $G$ is matching-decyclable if and only if $G$ has a spanning tree $T$ such that all leaves of $T$ are of degree at most 2 in $G$.

\end{proposition}

{\bf Proof.}

(a) Let $M$ be a decycling matching of $G$. Then, for every subgraph $H$ of $G$, $M\cap E(H)$ is a decycling matching of $H$.

(b) If $G$ is m-decyclable, the existence of a partition $(M,F)$ where $M$ is a matching and $F$ is acyclic implies $m = |M|+|F| \leq \lfloor\,\frac{n}{2}\,\rfloor + (n-1)=\lfloor\,\frac{3}{2}n\,\rfloor-1$.

(c) Trivial from the definition of sparse graph.

(d) Suppose that $G$ contains at most one vertex $v$ with $d_G(v)\leq 2$. Then  $m\geq (3(n-1)+2)/2>\lfloor\,\frac{3}{2}n\,\rfloor-1$, contradicting the definition of sparse graphs.

(e) Follows from the fact that the graphs $K_4$, $K_{3,3}$, and gem are bad.

(f) Let $G$ be a $2$-connected fairly cubic graph, and let $H$ be a subgraph of $G$. We claim that $H$ has at least two vertices of degree at most $2$. If $H = G$, then this is immediate by the definition of a fairly cubic graph. So we can assume that $H$ is a proper subgraph of $G$. If $H$ contains exactly one vertex $v$ with a neighbour outside $H$, then $v$ is a cut vertex of $G$, which contradicts the fact that $G$ is $2$-connected. Hence $H$ contains at least two vertices with a neighbour outside $H$, which proves our claim. This implies $|E(H)|\leq\lfloor\,\frac{3}{2} |V(H)|\,\rfloor-1$. Hence, $G$ is sparse.


(g) Assume that $G$ is not sparse. Then $G$ contains a bad subgraph $H$, i.e.,  $|E(H)|>\lfloor\,\frac{3}{2} |V(H)|\,\rfloor-1$. By item (b), this implies that $H$ is not m-decyclable. But this contradicts item (a). Therefore, item (g) follows.

(h) A decycling set of $K_{2,4}$ must contain at least three edges, but the size of a maximum matching in $K_{2,4}$ is two.

(i) The proof can be found in~\citep{lima16}.

(j) The proof can be found in~\citep{lima16}. \ $\Box$

\medskip

Since $K_{2,4}$ is sparse, Proposition~\ref{prop:super-prop}(h) implies that being sparse is not a sufficient condition for a graph to be m-decyclable. An interesting question is to find graph classes in which being m-decyclable is equivalent to being sparse. This question is dealt with in sections 4 and 5.

\if 10

\section{Sparse graphs}

\begin{lemma}\label{lem:pendant}
Let $v$ be a pendant vertex of a graph $G$. Then $G$ is sparse if and only if $G-v$ is sparse.
\end{lemma}
\vspace{-0.3cm}
\noindent {\bf Proof.} \ The `only if' part follows from Proposition~\ref{prop:sparse}. For the `if' part, let $H$ be a subgraph of $G$. If $v\not\in V(H)$ then $H$ is sparse. If $v\in V(H)$ then $H'=H-v$ is sparse and
$$m_{\scriptscriptstyle{H}}\leq m_{\scriptscriptstyle{H'}}+1\leq\lfloor\tfrac{3}{2}n_{\scriptscriptstyle{H'}}-1\rfloor +1=
\lfloor\tfrac{3}{2}n_{\scriptscriptstyle{H'}}\rfloor=\lfloor\tfrac{3}{2}(n_{\scriptscriptstyle{H}}-1)\rfloor\leq
\lfloor\tfrac{3}{2}n_{\scriptscriptstyle{H}}-1\rfloor,$$
i.e., $H$ is sparse. Hence, by Proposition~\ref{prop:sparse}, $G$ is sparse. \ $\Box$

\fi

\section{M-decyclable subcubic graphs}

In this section we study m-decyclable subcubic graphs. Let $\C$ be the class of 2-connected planar fairly cubic graphs. In~\citep{lima16} the authors show that a graph $G\in\C$ is m-decyclable if and only if $G$ has a Hamiltonian path whose endvertices are precisely the vertices of degree two in $G$. In fact, the assumptions ``2-connected'' and ``planar'' are not needed to state their result:

\begin{proposition}{\em ~\citep{lima16}}\label{prop:ham-path}
Let $G$ be a connected fairly cubic graph. Then $G$ is matching-decyclable if and only if there is a Hamiltonian path in $G$ whose endpoints are the vertices of degree two.
\end{proposition}

{\bf Proof.} If $G$ is matching-decyclable, by Proposition~\ref{prop:super-prop}(i) $G$ has a matching $M$ such that $G-M$ is a tree. Thus $|M|=|E(G)|-(n-1)= (\frac{3}{2}n-1)-(n-1)=\frac{n}{2}$, i.e., $M$ is a perfect matching. This implies that $G-M$ has $n-2$ vertices of degree two and two vertices $s$ and $t$ of degree one, i.e., it is a Hamiltonian path with endpoints $s$ and $t$. Since $d_G(s)=d_G(t)=2$, the first part follows. Conversely, if there is a Hamiltonian path $P$ in $G$ whose endpoints are the vertices of degree two, it is easy to see that the edges not in $P$ form a matching, i.e., $G$ is matching-decyclable. \ $\Box$

A simple by-product of the above proposition is the existence of a class of graphs in which being m-decyclable is equivalent to being Hamiltonian:

\begin{corollary}
Let $\C'=\{H\in\C:\mbox{the vertices of degree two in} \ H \ \mbox{are adjacent}\}$. Then $G\in\C'$ is m-decyclable if and only if $G$ is Hamiltonian.
\end{corollary}

As explained in~\citep{lima16}, for a graph $G\in\C$ the problem of deciding whether there is a Hamiltonian path whose endvertices are the vertices of degree two is NP-complete. Thus:

\begin{theorem}{\em \citep{lima16}}\label{lima}
Deciding whether a 2-connected planar fairly cubic graph is m-decyclable is NP-complete.
\end{theorem}

\begin{corollary}
Deciding whether a sparse graph is m-decyclable is NP-complete.
\end{corollary}

\noindent {\bf Proof.} \ Recall from Proposition~\ref{prop:super-prop}(f) that $2$-connected fairly cubic graphs are sparse. Thus, by Theorem~\ref{lima} deciding matching-decyclability is hard even for a subset of sparse graphs. \ $\Box$

Now we strength the result of Theorem~\ref{lima} in the following way. By the theorem, deciding matching-decyclability is NP-complete in the class $\D$ of 2-connected fairly cubic graphs. We show below that deciding matching-decyclability remains NP-complete in a proper subclass of $\D$, namely, Hamiltonian fairly cubic graphs. By Proposition~\ref{prop:ham-path}, we will show instead that, given a Hamiltonian fairly cubic graph $G$, the problem of deciding whether there is a Hamiltonian path in $G$ whose endpoints are the vertices of degree two is NP-complete. First, we need to consider the following problem:

{\sc Hamiltonian Cycle Containing a Specified Edge in a Cubic Graph}\\
{\em Input:} A cubic graph $H$, an edge $e$ of $H$.\\
{\em Question:} Does $H$ admit a Hamiltonian cycle containing edge $e$?

\begin{figure}[htbp]
\centering
\includegraphics[width=0.7\textwidth]{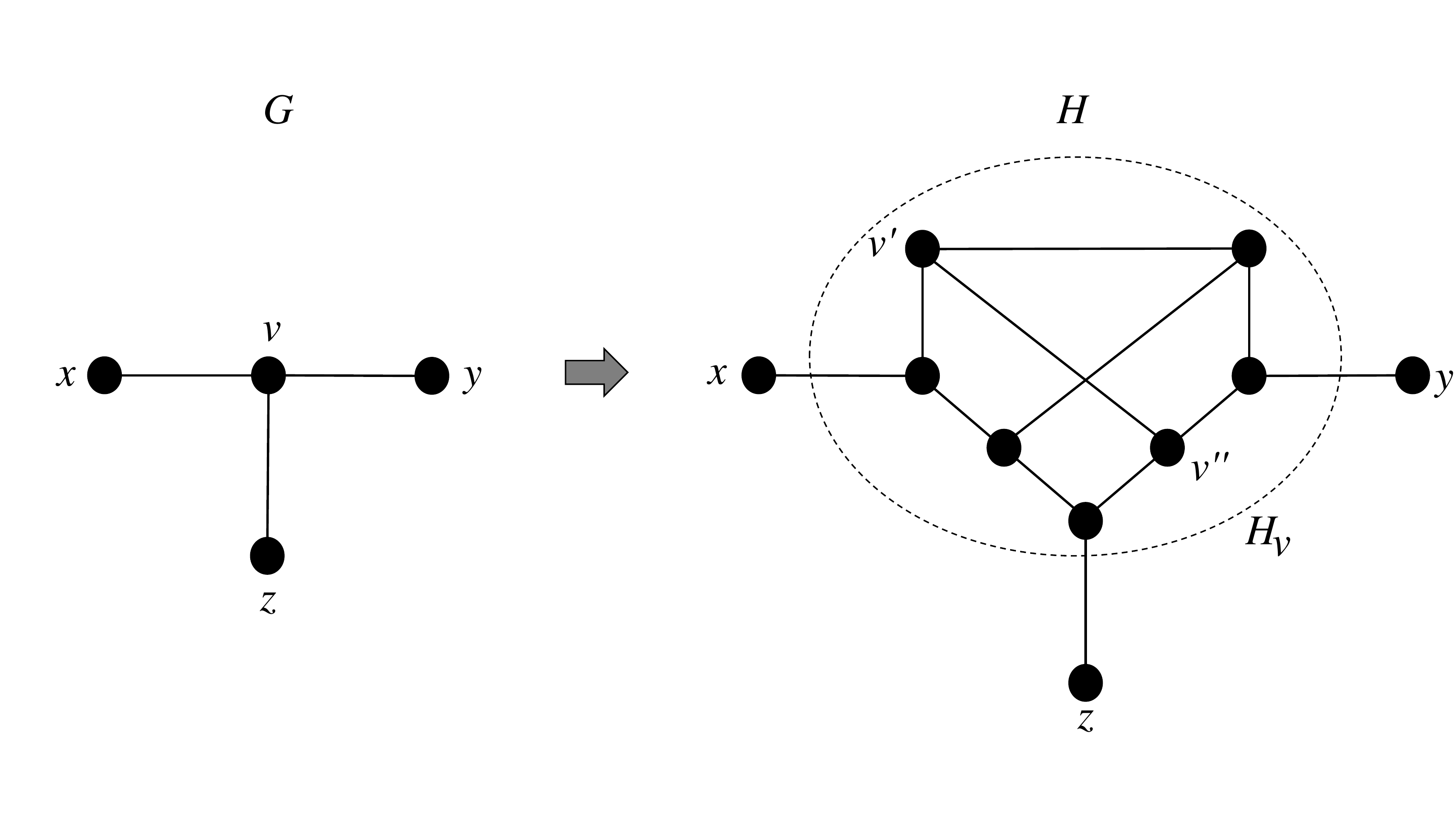}
\caption{Illustration for Lemma~\ref{lem:hc-edge}.}\label{fig:red-diamond}
\end{figure}

The above problem is easily seen to be in NP. The hardness proof is a straightforward reduction from the problem of checking whether a cubic graph $G$ is Hamiltonian~\citep{gjt76}. From $G$ we construct a cubic graph $H$ by replacing an arbitrarily chosen vertex $v$ of $G$ by the gadget $H_v$ depicted in Figure~\ref{fig:red-diamond}. In addition, we define $e=v'v''$. It is easy to see that $G$ admits a Hamiltonian cycle if and only if $H$ admits a Hamiltonian cycle containing edge $v'v''$. Therefore:


\begin{lemma}\label{lem:hc-edge}
The problem {\sc Hamiltonian Cycle Containing a Specified Edge in a Cubic Graph} is NP-complete.
\end{lemma}

\vspace{-0.5cm}

\begin{theorem}\label{thm:npc-cubic} Let $G$ be a Hamiltonian fairly cubic graph, and let $s,t\in V(G)$ such that $d_G(s)=d_G(t)=2$. Then deciding whether there is a Hamiltonian path in $G$ whose endpoints are $s$ and $t$ is NP-complete.
\end{theorem}

{\bf Proof.} The problem is clearly in NP, because given a path $P$ in $G$, one can easily check in polynomial time whether $P$ is Hamiltonian and has endpoints $s$ and $t$. The hardness proof uses a reduction from {\sc Hamiltonian Cycle Containing a Specified Edge in a Cubic Graph}. From an instance $(H,e)$ of this problem, we construct a Hamiltonian fairly cubic graph $G$ as follows. We can assume that $|V(H)|\geq 3$.

\begin{figure}[htbp]
\centering
\includegraphics[width=\textwidth]{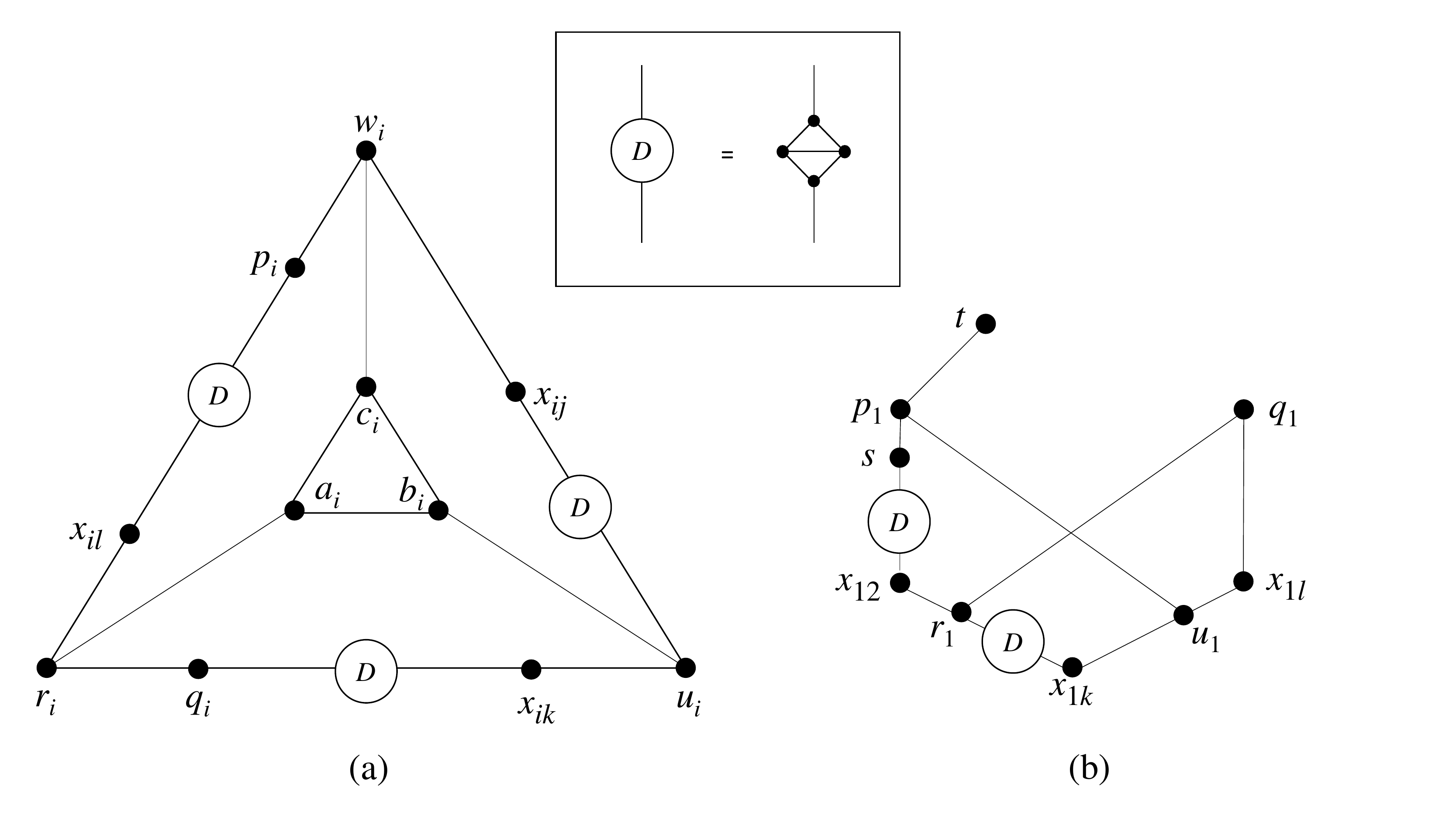}
\caption{Gadgets used in the reduction of Theorem~\ref{thm:npc-cubic}.}\label{fig:gadget-red2}
\end{figure}

{\bf Defining the gadgets.} Write $V(H)=\{v_1,v_2,\ldots,v_n\}$ $(n\geq 3)$, and assume without loss of generality that $e=v_1v_2$. We replace each vertex $v_i$, $2\leq i\leq n$, by the gadget $G_i$ depicted in Figure~\ref{fig:gadget-red2}(a). If $v_i$ has neighbors $v_j, v_k, v_{l}$ in $H$, then $G_i$ contains the vertices $x_{ij}, x_{ik}, x_{il}$ that will be used to connect $G_i$ to gadgets $G_j$, $G_k$, and $G_l$, respectively. We remark that their positions can be interchanged, i.e., in Figure~\ref{fig:gadget-red2}(a), $x_{ij}$ can occupy the position of $x_{ik}$ or $x_{il}$, etc.

Vertex $v_1$ is replaced by a different gadget $G_1$, shown in Figure~\ref{fig:gadget-red2}(b). The position of $x_{12}$ is fixed (between $D$ and $r_1$), and if $v_1$ has additional neighbors $v_k$ and $v_l$ in $H$ then $x_{1k}$ and $x_{1l}$ occupy the positions indicated in Figure~\ref{fig:gadget-red2}(b) (but their positions can also be switched, similarly as explained for Figure~\ref{fig:gadget-red2}(a)).

\begin{figure}[htbp]
\centering
\includegraphics[width=\textwidth]{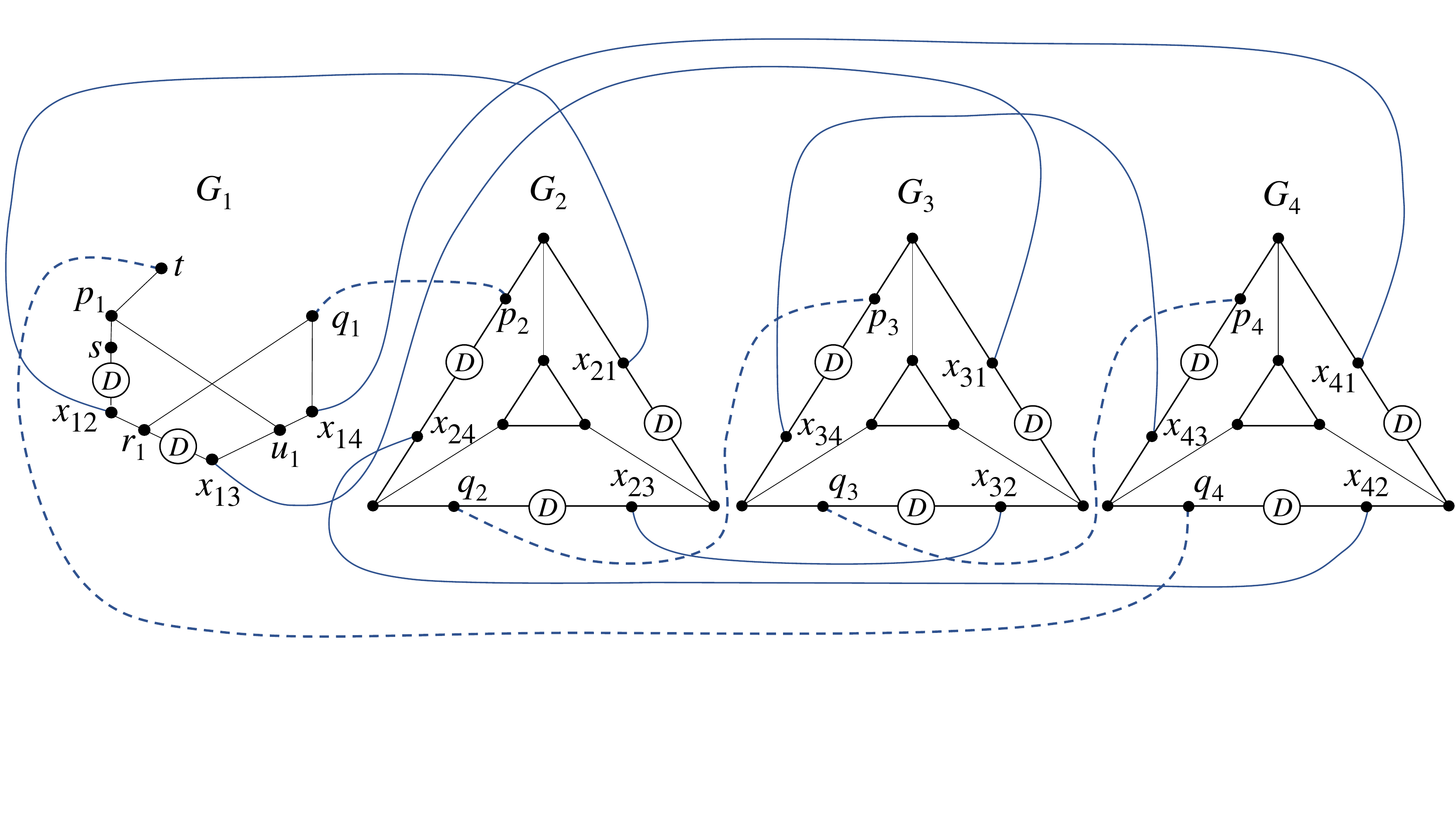}
\caption{Proof of Theorem~\ref{thm:npc-cubic}: construction of graph $G$ from $H=K_4$.}\label{fig:connect-gadgets2}
\end{figure}

{\bf Connecting the gadgets.} \ Figure~\ref{fig:connect-gadgets2} shows how to connect the gadgets. If $v_i$ has neighbors $v_j, v_k, v_{l}$ in $H$ then we link gadget $G_i$ to gadgets $G_j, G_k, G_{l}$ by creating the edges $x_{ij}x_{ji}, x_{ik}x_{ki}, x_{il}x_{li}$. Since $v_1$ and $v_2$ are neighbors, the edge $x_{12}\,x_{21}$ connecting gadgets $G_1$ and $G_2$ always exists. In addition, there are edges $q_1p_2, q_2p_3, q_3p_4, \ldots, q_{n-1}p_n$ and $q_n\,t$ (represented as dashed lines). Figure~\ref{fig:connect-gadgets2} shows the construction of $G$ from $H=K_4$.

{\bf Properties of $G$.} \ Note that $G$ is a fairly cubic graph, since vertices $s$ and $t$ have degree two, and the remaining vertices have degree three. Now, consider the following paths (where the symbol `$D$' represents a suitable subpath visiting all the vertices of a diamond) :
\[ P_1 = p_1 \ s \ D \ x_{12} \ r_1 \ D \ x_{1k} \ u_1 \ x_{1l} \ q_1 \ p_2,\]
\[ P_i = p_i \ D \ x_{il} \ r_i \ a_i \ b_i \ c_i \ w_i \ x_{ij} \ D \ u_i \ x_{ik} \ D \ q_i \ p_{i+1} \ \ (2\leq i\leq n-1),\]
\[P_n= p_n \ D \ x_{nl} \ r_n \ a_n \ b_n \ c_n \ w_n \ x_{nj} \ D \ u_n \ x_{nk} \ D \ q_n \ t \ p_1.\]

Note that the concatenation $P_1 \ P_2 \ \cdots \ P_{n-1} \ P_n$ is a Hamiltonian cycle. Thus, $G$ is a Hamiltonian fairly cubic graph, as required.

{\bf Properties of the gadgets.} \ We list below some important properties of the gadgets that will be useful for the proof. All of them can be easily checked by inspection.

{\em Property 1:} There is a (unique) Hamiltonian path $Q_i$ in $G_i$, $2\leq i\leq n$, with endpoints $x_{ij}$ and $x_{ik}$ (up to distinct ways of traversing the diamonds -- \ in fact, a diamond can be viewed as a vertex of degree two in all the properties listed in this subsection):
\[Q_i = x_{ij} \ D \ u_i \ b_i \ a_i \ c_i \ w_i \ p_i \ D \ x_{il} \ r_i \ q_i \ D \ x_{ik}.\]

{\em Property 2:} There is a (unique) Hamiltonian path $R_i$ in $G_i$, $2\leq i\leq n$, with endpoints $x_{ij}$ and $x_{il}$\;:
\[R_i = x_{ij} \ D \ u_i \ x_{ik} \ D \ q_i \ r_i \ a_i \ b_i \ c_i \ w_i \ p_i \ D \ x_{il}.\]

{\em Property 3:} There is a (unique) Hamiltonian path $S_i$ in $G_i$, $2\leq i\leq n$, with endpoints $x_{ik}$ and $x_{il}$\;:
\[S_i = x_{ik} \ D \ q_i \ r_i \ a_i \ c_i \ b_i \ u_i \ D \ x_{ij} \ w_i \ p_i \ D \ x_{il}.\]

{\em Property 4:} There is a (unique) Hamiltonian path $Z_i$ in $G_i$, $2\leq i\leq n$, with endpoints $p_i$ and $q_i$\;:
\[Z_i = p_i \ D \ x_{il} \ r_i \ a_i \ b_i \ c_i \ w_i \ x_{ij} \ D \ u_i \ x_{ik} \ D \ q_i.\]

{\em Property 5:} There is no Hamiltonian path in $G_i$, $2\leq i\leq n$, with an endpoint in the set $\{p_i, q_i\}$ and another endpoint in the set $\{x_{ij}, x_{ik}, x_{il}\}$.

To state the next property, we need some definitions. Let $T_i=\{p_i,q_i,x_{ij},x_{ik},x_{il}\}$, for $2\leq i\leq n$. Say that $T_i$ is the set of {\em terminals} of $G_i$. The vertices $x_{ij},x_{ik},x_{il}$ are the {\em type-$x$ terminals} of $T_i$. Similarly, we define  $T_1=\{t, q_1,x_{12},x_{1k},x_{1l}\}$ as the set of terminals of $G_1$, where $x_{12}$, $x_{1k}$, and $x_{1l}$ are the type-$x$ terminals of $G_1$.

{\em Property 6:} For $2\leq i\leq n$, there is no partition of $V(G_i)$ into two subsets $X_i$ and $Y_i$ such that both $X_i$ and $Y_i$ form nontrivial paths starting and ending at terminals. In other words, it is not possible to cover all the vertices of $G_i$ using two nontrivial disjoint paths whose endpoints are terminals.



The diamonds in Figures~\ref{fig:gadget-red2}(a) and~\ref{fig:gadget-red2}(b) have the purpose of forcing the paths to visit some parts of the gadgets. Vertices of degree two could play the same role as the diamonds, but since we need to construct a fairly cubic graph $G$, the use of diamonds is simply an artifice to make all the vertices have degree three (except $s$ and $t$, of course).

{\bf Completing the proof.} Let us prove that there is a Hamiltonian cycle in $H$ containing edge $e=v_1v_2$ if and only if there is a Hamiltonian path from $s$ to $t$ in $G$.

Suppose first that there is a Hamiltonian cycle $C$ in $H$ containing edge $e=v_1v_2$. Suppose without loss of generality that $C=v_1 v_2 v_3 \ldots v_{n-1} v_n v_1$. For $h=2,3,\ldots,n$, let $P(x_{h\,h-1},x_{h\,h+1})$ be a Hamiltonian path of $G_h$ from $x_{h\,h-1}$ to $x_{h\,h+1}$ (by Properties 1, 2, and 3 such a path is one of $Q_h, R_h, S_h$). We remark that $h+1\equiv 1$ when $h=n$.

The following path is a Hamiltonian path in $G$ from $s$ to $t$:
\[s D x_{_{12}} P(x_{_{21}},x_{_{23}}) x_{_{23}} x_{_{32}} P(x_{_{32}},x_{_{34}}) x_{_{34}} x_{_{43}} \ldots x_{_{n-1\,n}} x_{_{n\,n-1}} P(x_{_{n\,n-1}},x_{_{n1}}) x_{_{n1}} x_{_{1n}} q_{_1} r_{_1} D x_{_{1k}} u_{_1} p_{_1} t.\]

(The above path assumes that $x_{1n}=x_{1l}$ in Figure~\ref{fig:gadget-red2}(b). If $x_{1n}=x_{1k}$, then the final part of the Hamiltonian path is:
\[\ldots \ x_{_{1n}}\,D\,r_{_1}\,q_{_1}\,x_{_{1l}}\,u_{_1}\,p_{_1}\,t.\]

This concludes the first part of the proof. Suppose now that there is a Hamiltonian path $P_{st}$ from $s$ to $t$ in $G$. We need the following definition. A {\em visit to a gadget} $G_i$, $1\leq i\leq n$, is a maximal subpath $P'$ of $P_{st}$ such that $P'$ contains only vertices of $G_i$. Note that a visit to a gadget $G_i$ is a path in $G_i$ starting and ending at terminals of $G_i$. Since each $G_i$, $2\leq i\leq n$, contains five terminals, it is visited at most twice. The claim below says that $G_2, G_2, \ldots, G_n$ cannot be visited twice. ($G_1$ is an exception to this rule.)

{\em Claim 1:} In a Hamiltonian path $P_{st}$ from $s$ to $t$ in $G$, each gadget $G_i$, $2\leq i\leq n$, is visited exactly once.

{\em Proof of Claim 1:} Suppose by contradiction that some $G_i$, $2\leq i\leq n$, is visited twice. Let $P'$ and $P''$ be the two paths representing such visits. Then $V(P')$ and $V(P'')$ is a partition of $V(G_i)$ such that both $P'$ and $P''$ are paths starting and ending at terminals. But this contradicts Property 6. Hence, the claim follows. \ $\Box$

The path $P_{st}$ starts at gadget $G_1$, and ends at the same gadget, so $G_1$ is visited more than once. Consider the first visit to $G_1$, that starts at vertex $s$.

$\blacktriangleright$ If the first visit to $G_1$ leaves it via $q_1$ then we have the following four possibilities:
\[ s \ p_1 \ u_1 \ x_{1l} \ q_1, \ \ \
 s \ p_1 \ u_1 \ x_{1k} \ D \ r_1 \ q_1, \ \ \
 s \ D \ x_{12} \ r_1 \ q_1, \ \ \
 s \ D \ x_{12} \ r_1 \ D \ x_{1k} \ u_1 \ x_{1l} \ q_1.\]

In all the possibilities above, it is easy to see that, besides $t$, at least one more vertex of $G_1$ is not visited before leaving the gadget. The path $P_{st}$ then follows edge $q_1p_2$ and enters $G_2$. By Claim 1, $G_2$ must be visited only once. Thus the visit to $G_2$ must pass through all of its vertices. But Property 5 tells us that the visit to $G_2$ must end precisely at $q_2$; in addition, Property 4 tells us that that the visit to $G_2$ is the path $Z_2$. This process continues, and each new visit to a gadget $G_i$, $2\leq i\leq n$, by Claim 1 and Properties 4-5, consists precisely of the path $Z_i$. Eventually, there is a visit to a gadget $G_j$, which is left via the edge $q_j t$, concluding the traversal of $G$. But since at least one vertex of $G_1$ has not been visited, this contradicts the fact that $P_{st}$ is a Hamiltonian path. Hence:

{\em Claim 2:} The first visit to $G_1$ cannot leave it via $q_1$.

In fact, the above arguments show that if a visit to $G_1$ (not necessarily the first one) leaves it via $q_1$ then the path $P_{st}$ returns to $G_1$ using an edge $q_j t$. Hence:

{\em Claim 3:} If a visit to $G_1$ leaves it via $q_1$ then the only vertex of $G_1$ to be visited subsequently is $t$.

$\blacktriangleright$ The preceding discussion leads to the conclusion that the first visit to $G_1$ leaves it via a type-$x$ terminal. Therefore, $P_{st}$ must then enter a gadget $G_i$, $2\leq i\leq n$, at a type-$x$ terminal as well. By Claim 1, all the vertices of $G_i$ must be visited before leaving it, and by Properties 1 to 3 the visit to $G_i$ consists of one of the paths $Q_i, R_i, S_i$. This implies that the next visit to a gadget $G_{i'}$, $i'\neq i$, similarly consists of one of the paths $Q_{i'}, R_{i'}, S_{i'}$. The process continues, and eventually there is a visit to a gadget $G_j$ which is left via a type-$x$ terminal $x_{j1}$, and the path returns to gadget $G_1$ at one of its type-$x$ terminals. Hence:

{\em Claim 4:} The first visit to $G_1$ leaves it via a type-$x$ terminal, and $P_{st}$ returns to $G_1$ at another of its type-$x$ terminals.

{\em Claim 5:} If a visit to $G_1$ (not necessarily the first one) leaves it via a a type-$x$ terminal then $P_{st}$ returns to $G_1$ at another of its type-$x$ terminals.

We now need to analyze the possible ways the path $P_{st}$ leaves and returns to $G_1$.

$\blacktriangleright\blacktriangleright$ Suppose that the first visit to $G_1$ passes by all of its type-$x$ terminals. The possibilities are:

\[s \ D \ x_{12} \ r_1 \ D \ x_{1k} \ u_1 \ x_{1l} \ \ \  \mathrm{and} \ \ \
s \ D \ x_{12} \ r_1 \ q_1 \ x_{1l} \ u_1 \ x_{1k}.\]

By Claim 5, $P_{st}$ must return to $G_1$ at one of its type-$x$ terminals, but all of them have been already visited. Thus the first visit to $G_1$ cannot pass by all of its type-$x$ terminals.

$\blacktriangleright\blacktriangleright$ Suppose now that the first visit to $G_1$ passes by exactly two of its type-$x$ terminals. The table below lists the possibilities. A symbol ``?'' means that the traversal cannot continue (thus the corresponding possibility is impossible).

\begin{table}[htb]
\centering
\small
\begin{tabular}{ |c|c|c|c| }
\toprule
1st visit to $G_1$ & 2nd visit to $G_1$ & 3rd visit to $G_1$ & Observations\\ \hline
$s \ p_1 \ u_1 \ x_{1l} \ q_1 \ r_1 \ D \ x_{1k}$ & $x_{12} \ D \ ?$ & -- & impossible \\
$s \ p_1 \ u_1 \ x_{1l} \ q_1 \ r_1 \ x_{12}$ & $x_{1k} \ D \ ?$& -- & impossible \\
$s \ p_1 \ u_1 \ x_{1k} \ D \ r_1 \ x_{12}$ & $x_{1l} \ q_1$ & $t$ \ (Claim 3) & a $D$ is not visited \\
$s \ p_1 \ u_1 \ x_{1k} \ D \ r_1 \ q_1 \ x_{1l}$ & $x_{12} \ D \ ?$ & -- & impossible \\
$s \ D \ x_{12} \ r_1 \ D \ x_{1k}$ & $x_{1l} \ q_1$ & $t$ \ (Claim 3) & $u_1$ and $p_1$ are not visited \\
$s \ D \ x_{12} \ r_1 \ D \ x_{1k}$ & $x_{1l} \ u_1 \ p_1 \ t$ & -- & $q_1$ is not visited \\
$s \ D \ x_{12} \ r_1 \ q_1 \ x_{1l}$ & $x_{1k} \ u_1 \ p_1 \ t$ & -- & a $D$ is not visited \\
\bottomrule
\end{tabular}
\caption{Possibilities for the case when the first visit to $G_1$ passes by exactly two of its type-$x$ terminals.\label{tab:two-term}}
\end{table}

By Table~\ref{tab:two-term}, the first visit to $G_1$ cannot pass by exactly two of its type-$x$ terminals.

$\blacktriangleright\blacktriangleright$ Finally, suppose that the first visit to $G_1$ passes by exactly one of its type-$x$ terminals. The table below lists the possibilities.

\begin{table}[!h]
\centering
\small
\begin{tabular}{ |c|c|c|c| }
\toprule
1st visit to $G_1$ & 2nd visit to $G_1$ & 3rd visit to $G_1$ & Observations\\ \hline
$s \ p_1 \ u_1 \ x_{1l}$ & $x_{1k} \ D \ r_1 \ x_{12}$ & ? & impossible: all type-$x$ terminals already visited\\
& $x_{1k} \ D \ r_1 \ q_1$ & $t$ \ (Claim 3) & a $D$ and $x_{12}$ are not visited \\
& $x_{12} \ r_1 \ D \ x_{1k}$ & ? & impossible: all type-$x$ terminals already visited\\
& $x_{12} \ r_1 \ q_1$ & $t$ \ (Claim 3) & the $D$'s and $x_{1k}$ are not visited\\ \hline
$s \ p_1 \ u_1 \ x_{1k}$ & $x_{1l} \ q_1$ & $t$ \ (Claim 3) & the $D$'s, $x_{12}$, and $r_1$ are not visited \\
& $x_{1l} \ q_1 \ r_1 \ x_{12}$ & ? & impossible: all type-$x$ terminals already visited\\
& $x_{12} \ r_1 \ q_1$ & $t$ \ (Claim 3) & the $D$'s and $x_{1l}$ are not visited\\
& $x_{12} \ r_1 \ q_1 \ x_{1l}$ & ? & impossible: all type-$x$ terminals already visited\\ \hline
$s \ D \ x_{12}$ & $x_{1k} \ u_1 \ x_{1l}$ & ? & impossible: all type-$x$ terminals already visited \\
& $x_{1k} \ u_1 \ x_{1l} \ q_1$ & $t$ & a $D$, $r_1$, and $p_1$ are not visited \\
& $x_{1k} \ u_1 \ p_1 \ t$ & -- &  a $D$, $r_1$, $q_1$, and $x_{1l}$ are not visited \\
& $x_{1k} \ D \ r_1 \ q_1$ & $t$ & $u_1$, $p_1$, and $x_{1l}$ are not visited \\
& $x_{1k} \ D \ r_1 \ q_1 \ x_{1l}$ & ? & impossible: all type-$x$ terminals already visited \\
& $x_{1k} \ D \ r_1 \ q_1 \ x_{1l} \ u_1 \ p_1 \ t$ & -- & {\bf ok: all vertices are visited}\\
& $x_{1l} \ u_1 \ x_{1k}$ & ? & impossible: all type-$x$ terminals already visited \\
& $x_{1l} \ u_1 \ x_{1k} \ D \ r_1 \ q_1$ & $t$ & $p_1$ is not visited \\
& $x_{1l} \ u_1 \ p_1 \ t$ & -- & a $D$, $r_1$, $q_1$, and $x_{1k}$ are not visited \\
& $x_{1l} \ q_1$ & $t$ & a $D$, $r_1$, $u_1$, $p_1$, and $x_{1k}$ are not visited\\
& $x_{1l} \ q_1 \ r_1 \ D \ x_{1k}$ & ? & impossible: all type-$x$ terminals already visited \\
& $x_{1l} \ q_1 \ r_1 \ D \ x_{1k} \ u_1 \ p_1 \ t$ & -- & {\bf ok: all vertices are visited} \\
\bottomrule
\end{tabular}
\caption{Possibilities for the case when the first visit to $G_1$ passes by exactly one of its type-$x$ terminals.\label{tab:one-term}}
\end{table}

From Table~\ref{tab:one-term}, we conclude that $P_{st}$ visits $G_1$ twice: the first visit to $G_1$ is
\[s \ D \ x_{12} \,\]
and the second visit is
\[x_{1k} \, D \, r_1 \, q_1 \, x_{1l} \, u_1 \, p_1 \, t \ \ \ \ \mbox{or} \ \ \ \ x_{1l} \, q_1 \, r_1 \, D \, x_{1k} \, u_1 \, p_1 \, t.\]
After the first visit to $G_1$, the path $P_{st}$ enters $G_2$ via $x_{21}$.
By Claim 1, all the vertices of $G_2$ must be visited before leaving it, and by Properties 1 to 3 the visit to $G_2$ consists of one of the paths $Q_2, R_2, S_2$. Note that $P_{st}$ can only return to $G_1$ after visiting all the gadgets $G_i$, $2\leq i\leq n$, because the second visit to $G_1$ ends at $t$, the final destination of $P_{st}$. Thus, after leaving $G_2$ via one of its type-$x$ terminals, the next visit is made to a gadget $G_{i}$, $i\notin\{1,2\}$, that similarly consists of one of the paths $Q_{i}, R_{i}, S_{i}$. The process continues, and eventually there is a visit to a gadget $G_j$ which is left via a type-$x$ terminal $x_{j1}$, and the path finally returns to $G_1$ to make the second visit to it. Assume without loss of generality that $P_{st}$ visits $G_1, G_2, G_3, \ldots, G_n, G_1$ in this order. This corresponds to a Hamiltonian cycle $v_1,v_2,v_3,\ldots,v_n,v_1$ in $H$ that contains the edge $e=v_1v_2$. This concludes the proof. \ $\Box$

\section{M-decyclable chordal graphs}

\begin{figure}[htbp]
\centering
\includegraphics[width=0.9\textwidth]{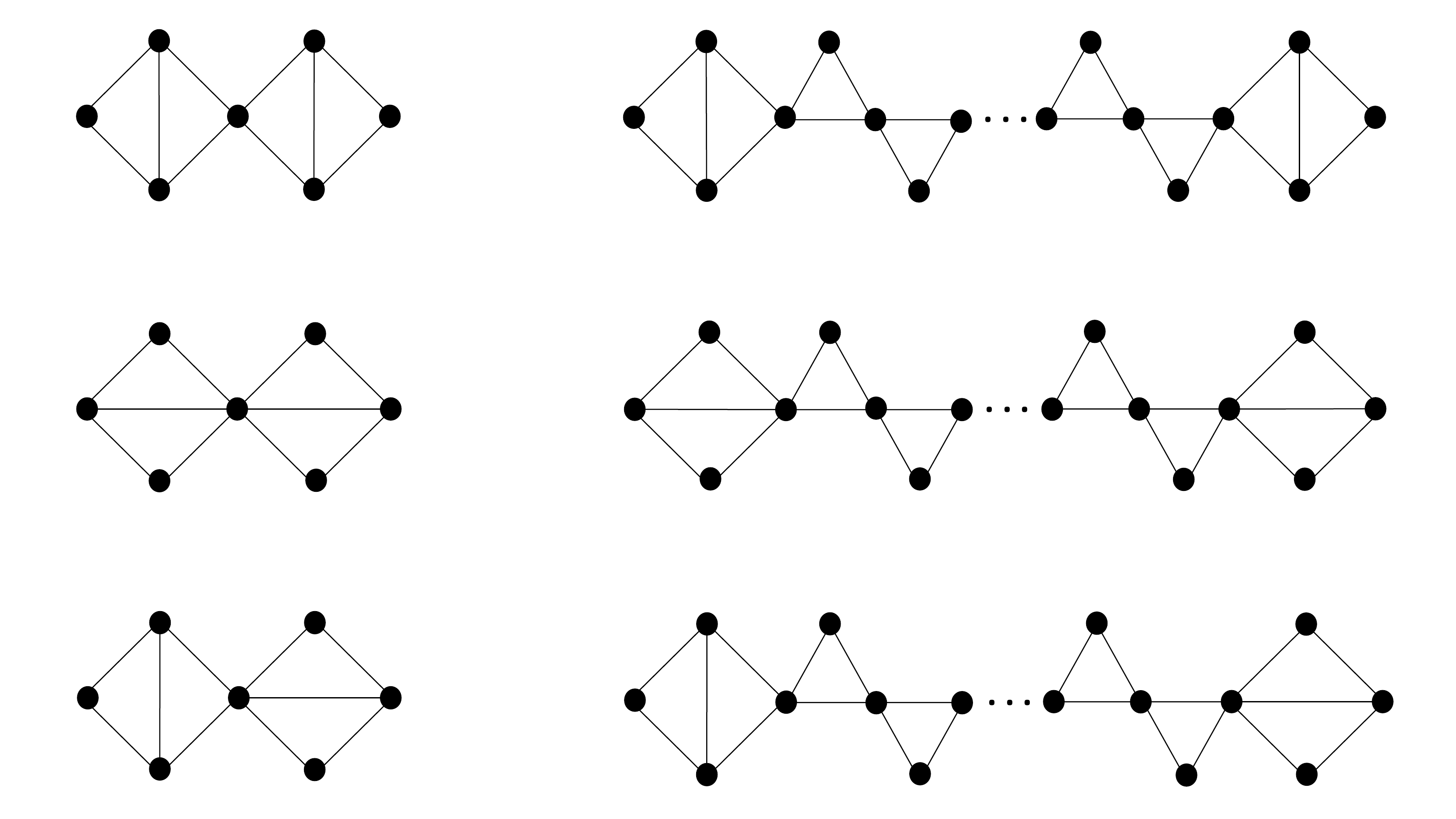}
\caption{Some examples of Chains.}\label{fig:chain}
\end{figure}

A {\em chain} is a graph containing exactly two leaf blocks, such that: $(i)$ if $B$ is a leaf block then $B$ is a diamond; \ $(ii)$ if $B$ is not a leaf block then $B$ is a triangle. See Figure~\ref{fig:chain}.

\begin{theorem}\label{thm:mdecchordal}
Let $G$ be a chordal graph. Then the following are equivalent:
\begin{itemize}
\setlength\itemsep{0em}
\item[$(a)$] $G$ is m-decyclable;
\item[$(b)$] $G$ is sparse;
\item[$(c)$] $G$ contains no chain, and each block of $G$ is a diamond, a triangle, or a bridge.
\end{itemize}
\end{theorem}

{\bf Proof.}

$(a)\Rightarrow (b)$. Follows from Proposition~\ref{prop:super-prop}(g).

\begin{figure}[!h]
\centering
\includegraphics[width=0.4\textwidth]{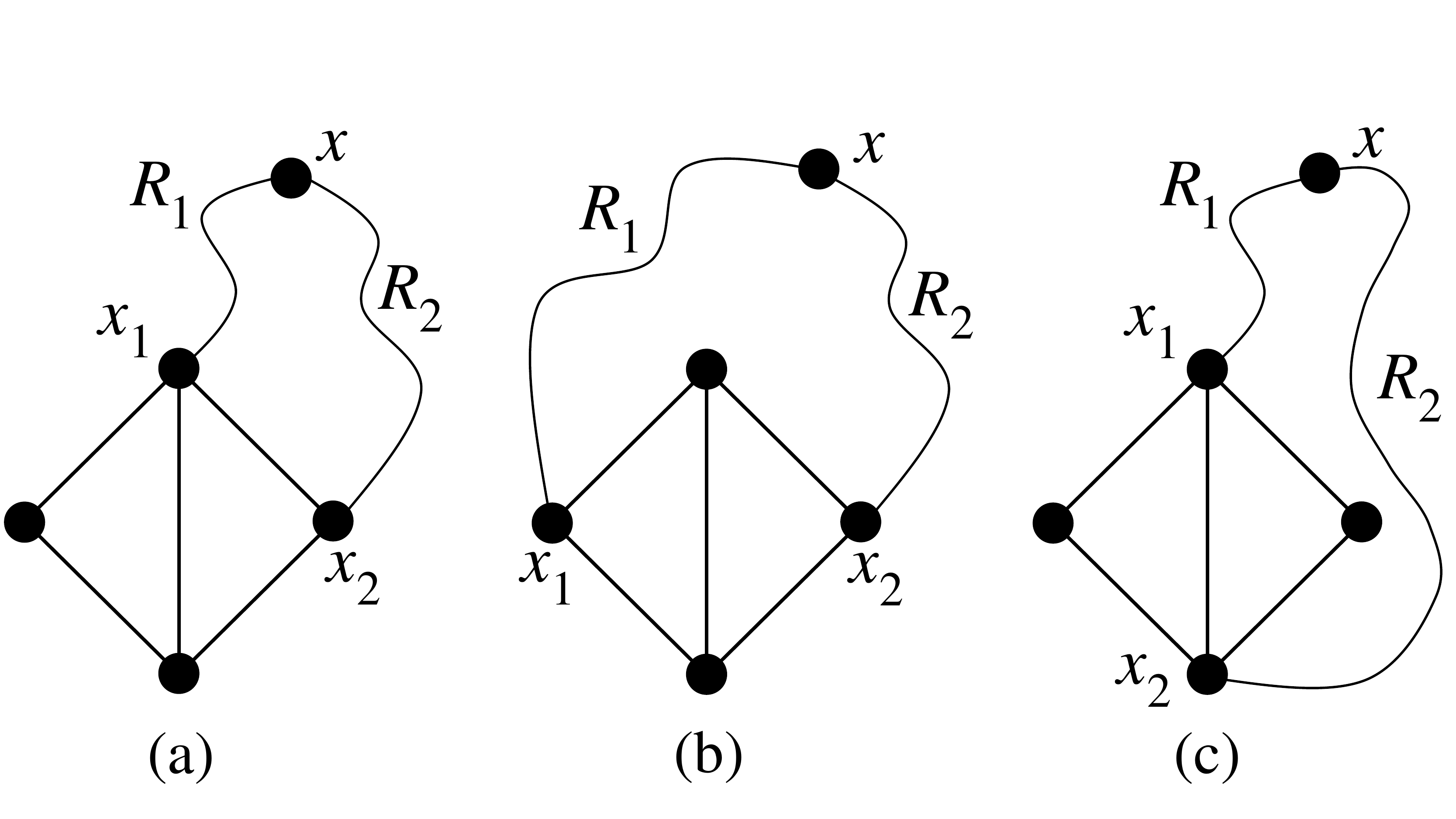}
\caption{Case 1 in the proof of $(b) \Rightarrow (c)$, Theorem~\ref{thm:mdecchordal}.\label{fig:x1x2}}
\end{figure}

\begin{figure}[htbp]
\centering
\includegraphics[width=0.45\textwidth]{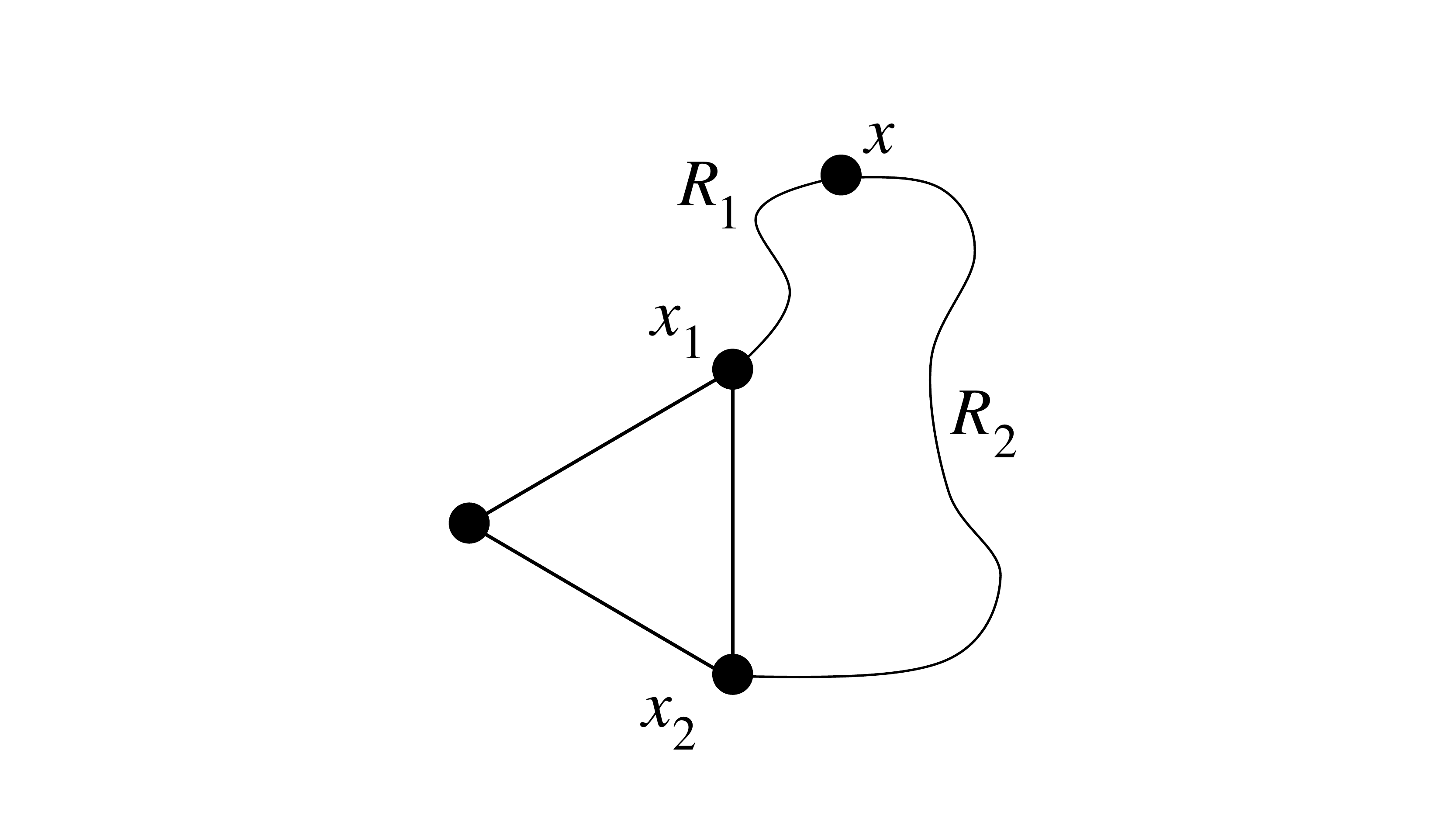}
\caption{Case 2 in the proof of $(b) \Rightarrow (c)$, Theorem~\ref{thm:mdecchordal}.\label{fig:x1x2b}}
\end{figure}

$(b)\Rightarrow (c)$. If $G$ is sparse then, by Proposition~\ref{prop:super-prop}(e), $G$ contains no induced gem. Therefore, by Theorem 2.5 in~\citep{howorka81}, every $k$-cycle in $G$ has at least $\lfloor\,\frac{3}{2}(k-3)\,\rfloor$ chords. But note that if $k\geq 5$ then such a cycle (together with its chords) forms a bad subgraph of $G$, a contradiction. This implies that $G$ contains no $k$-cycles for $k\geq 5$. Since $G$ is chordal and, by Proposition~\ref{prop:super-prop}(e), $G$ contains no $K_4$, the only possible cycles in $G$ are triangles and $4$-cycles having exactly one chord (diamonds). Now, consider a block $B$ of $G$ that is not a bridge, and two vertices $a_1,a_2\in V(B)$. Let $P$ and $Q$ be two internally disjoint paths linking $a_1$ and $a_2$ in $B$. By the preceding discussion, the graph $B'$ induced by $V(P)\cup V(Q)$ is either a diamond or a triangle. We analyze two cases:
\begin{enumerate}
\item $B'$ is a diamond. If $V(B)\setminus V(B')\neq\emptyset$, consider a path $R$ leaving $B'$ at $x_1$ and returning to $B'$ at $x_2\neq x_1$ such that $R$ visits $x\in V(B)\setminus V(B')$. Let $R_1$ and $R_2$ be the subpaths of $R$ from $x_1$ to $x$ and from $x$ to $x_2$, respectively. Figure~\ref{fig:x1x2} shows the three possible cases for $x_1$ and $x_2$. Figures~\ref{fig:x1x2}(a) and~\ref{fig:x1x2}(b) contain cycles of size at least five, a contradiction. In Figure~\ref{fig:x1x2}(c), both $R_1$ and $R_2$ must consist of a single edge each ($R_1=x_1x$ and $R_2=xx_2$), in order to avoid the existence of a cycle of size at least five in $G$; but then the subgraph of $G$ induced by $E(B')\cup\{x_1x,xx_2\}$ is bad (contains $5$ vertices and $7$ edges), another contradiction. Hence, $V(B)\setminus V(B')=\emptyset$, i.e., $B=B'$.

\item $B'$ is a triangle. If $B=B'$, we are done. Otherwise, as in the previous case, we can similarly define $x_1,x_2,x,R_1,R_2$ (see Figure~\ref{fig:x1x2b}). Note that both $R_1$ and $R_2$ must consist of a single edge each, in order to avoid a cycle of size greater than four
    in $G$. Hence, $B$ contains a diamond $B''$ induced by $E(B')\cup\{x_1x,xx_2\}$. By Case 1, no vertices outside $V(B'')$ are possible; therefore, $B=B''$.
\end{enumerate}

From the above cases we conclude that each block of $G$ is a diamond, a triangle, or a bridge. To prove that $G$ contains no chain as a subgraph, note that a chain with $k$ triangle blocks ($k\geq 0$) has $2k+7$ vertices and $3k+10$ edges, i.e., it is a bad subgraph; by Proposition~\ref{prop:super-prop}(c), this concludes the proof of $(b)\Rightarrow (c)$.

$(c)\Rightarrow (a)$. Let $G'$ be the graph obtained by the removal of the bridges of $G$. As $G'$ has no chain then a decycling matching $M$ of $G'$ (and thus of $G$) can be trivially obtained as follows: (i) unmark all the edges; (ii) include in $M$ two disjoint edges for each diamond block of $G'$; (iii) mark every edge belonging to $M$ or incident with some edge in $M$; (iv) include in $M$ one non-marked edge $e_T$ for each triangle block $T$ of $G'$ so that the chosen edges are pairwise disjoint. \ $\Box$

\begin{corollary}\label{cor:mdecchordal}
Let $G$ be a $2$-connected chordal graph. Then $G$ is m-decyclable if and only if $G$ is a diamond or a triangle.
\end{corollary}

The following result refines the polynomial-time algorithm presented in~\citep{lima16}:

\begin{theorem}\label{thm:recog-mdecchordal}
M-decyclable chordal graphs can be recognized in $O(n)$ time.
\end{theorem}

{\bf Proof.} \ Given a chordal graph $G$, first check whether $m\leq\lfloor\,\frac{3}{2}n\,\rfloor-1$. If so, we have $m=O(n)$, and then the block decomposition of $G$ can be obtained in $O(n)$ time using standard depth-first search. Next, check whether every block of the decomposition is a diamond, a triangle, or a bridge (this can be easily done in $O(1)$ time per block: if a block has more than four vertices then the process stops, otherwise if it has four vertices then it must have exactly five edges). Finally, remove the bridges and check whether each resulting connected component contains at most one diamond block. The entire process clearly runs in $O(n)$ time. In addition, if each resulting connected component in the above procedure indeed contains at most one diamond block then a decycling matching $M$ of $G$ can be trivially obtained in $O(n)$ as follows: (i) unmark the current edges of $G$; (ii) include in $M$ two disjoint edges for each diamond block of $G'$; (iii) mark every edge belonging to $M$ or incident with some edge in $M$; (iv) include in $M$ one non-marked edge $e_T$ for each triangle block $T$ of $G'$ so that the chosen edges are pairwise disjoint. \ $\Box$

\begin{corollary}
If $M$ is a decycling matching of a chordal graph $G$ then $M$ contains exactly $2d+t$ edges, where $d$ is the number of diamond blocks and $t$ the number of triangle blocks of $G$.
\end{corollary}

{\bf Proof.} \ Follows directly from the previous proof. \ $\Box$

\medskip

A {\em star} is a graph isomorphic to $K_{1,p}$, for a natural number $p$. The {\em center} of the star is the vertex of degree $p$. A {\em double star} is the union of two stars together with an edge joining their centers. A {\em ``triangle with pendant vertices''} is a graph containing a triangle formed by vertices $a,b,c$, such that each remaining vertex $v$ (if any) has exactly one neighbor in $\{a,b,c\}$. A {\em ``diamond with pendant vertices all attached to a same triangle''} is a graph containing a diamond formed by vertices $a,b,c,d$ and edges $ab, ac, ad, bc, cd$, such that each remaining vertex $v$ (if any) has exactly one neighbor in  $\{a,b,c\}$.

By combining the fact that a split graph contains no $2K_2$ as an induced subgraph (see~\citep{fh77}) with item (c) in Theorem~\ref{thm:mdecchordal}, an easy consequence of Theorem~\ref{thm:mdecchordal} is:

\begin{corollary}\label{coro:split}
Let $G$ be a connected split graph. Then $G$ is matching-decy\-cla\-ble if and only if $G$ is a star, a double star, a triangle with pendant vertices, or a diamond with pendant vertices all attached to a same triangle.
\end{corollary}


\section{M-decyclable distance-hereditary graphs}

In this section we present a characterization of distance-hereditary graphs that are m-decyclable. The arguments used in the proofs are strongly based on sparseness, and extend in some sense those used in the previous section for chordal graphs. As we shall see, the concept of {\em ear decomposition}~\citep{whitney} will be useful for the proofs.



We use the following notation. Let $C = v_1v_2\ldots v_kv_1$ be a $k$-cycle in $G$. The subgraph of $G$ induced by $V(C)$ is denoted by $G_C$. A $j${\it -chord} in $C$ \mbox{$(2\leq j\leq\lfloor\frac{k}{2}\rfloor)$} is an edge joining two vertices $v_i,v_{i+j}\in V(C)$ (where $i+j$ is taken modulo $k$). Note that $K_{3,3}^-$ consists of a $6$-cycle plus two $3$-chords.

\begin{lemma}\label{lem:possible-cycles}
Let $G$ be a sparse distance-hereditary graph, and let $C$ be a cycle of $G$. Then $G_C$ is one of the following graphs: triangle, square, diamond, or $K_{3,3}^-$.
\end{lemma}

{\bf Proof.} \ Let $C$ be a $k$-cycle of $G$. If $k=3$ then $G_C$ is a triangle, and if $k=4$ then, by Proposition~\ref{prop:super-prop}(e), $G_C$ is either a square or a diamond.

Assume $k=5$. Since $G$ is distance-hereditary, $G_C$ is neither a hole nor a house. Thus $C$ must contain at least two chords. But this implies that $G_C$ is a bad subgraph of $G$ with five vertices and at least seven edges, a contradiction. Hence, $G$ contains no $5$-cycles.

Assume now $k=6$. Since $G$ is distance-hereditary, $G_C$ is neither a hole nor a domino. In addition, $C$ contains no $2$-chord, for otherwise $G$ would contain a $5$-cycle, and this is impossible by the previous paragraph. Since by Proposition~\ref{prop:super-prop}(c) $G_C$ is sparse, $C$ cannot contain three or more chords. Thus $G_C$ is the graph $K_{3,3}^-$.

In all the remaining cases, $C$ must contain at least one chord, because $G$ contains no holes. The cases are explained below.

$\blacktriangleright \ k=7$: If $C$ contains a $2$-chord then $G_C$ contains a $6$-cycle $C'$; but, by the previous analysis, $G_{C'}$ is the graph $K_{3,3}^-$, implying the existence of two additional chords in $C$ and therefore at least ten edges in $G_C$, a contradiction. If $C$ contains a $3$-chord then $G_C$ contains a $5$-cycle, which we have already seen to be impossible. Hence, $G$ contains no $7$-cycles.

$\blacktriangleright \ k=8$: If $C$ contains a $2$-chord then $G_C$ contains a $7$-cycle, which is impossible by the previous case. If $C$ contains a $3$-chord, say $v_1v_4$, then $G_C$ contains a $6$-cycle $C'=v_1v_4v_5v_6v_7v_8v_1$ such that $G_{C'}$ is again the graph $K_{3,3}^-$; now $C'$ contains an additional pair of $3$-chords, $\{v_1v_6,v_4v_7\}$ or $\{v_4v_7,v_5v_8\}$ or $\{v_1v_6,v_5v_8\}$, but in any case additional $6$-cycles other than $C'$ exist in $G_C$, each of them requiring one additional chord not yet listed; and this implies the existence of more than eleven edges in $G_C$, a contradiction. Finally, if $C$ contains a $4$-chord then $G_C$ contains two $5$-cycles, which is impossible. Hence, $G$ contains no $8$-cycles.

$\blacktriangleright \ k=9$: If $C$ contains a $2$-chord (resp., $3$-chord, $4$-chord) then $G_C$ contains a $8$-cycle (resp., $7$-cycle, $5$-cycle), which is impossible by the previous cases. Hence, $G$ contains no $9$-cycles.

$\blacktriangleright \ k=10$: For any $j\in\{2,3,4\}$, if $C$ contains a $j$-chord then $G_C$ contains an $(11-j)$\,-\,cycle, which is impossible by the previous cases. If $C$ contains a $5$-chord then $G_C$ contains two distinct $6$-cycles, each requiring two additional chords; this implies the existence of at least fifteen edges in $G_C$, a contradiction. Hence, $G$ contains no $10$-cycles.

$\blacktriangleright \ k\geq 11$: In this case, the existence of any chord in $C$ implies a $k'$-cycle in $G_C$ for $7 \leq k'< k$, and this is impossible by the previous cases. Hence, $G$ contains no $k$-cycles for $k\geq 11$. \ $\Box$

\medskip

The following definitions are necessary for the next theorem. The {\em union} of two graphs $G_1$ and $G_2$ is the graph $G_1\cup G_2$ such that $V(G_1\cup G_2)=V(G_1)\cup V(G_2)$ and $E(G_1\cup G_2)=E(G_1)\cup E(G_2)$. An {\em ear} of a graph $G$ is a maximal path $P$ whose internal vertices have degree two in $G$, and whose endpoints have degree at least three in $G$. An {\em ear decomposition} of a graph $G$ is a decomposition $G_0\cup G_1\cup\cdots\cup G_p$ of $G$ such that $G_0$ is a cycle and $G_i$ is an ear of $G_0\cup G_1\cup\cdots\cup G_i$. It is well known that a graph is $2$-connected if and only if it admits an ear decomposition. Furthermore, every cycle in a $2$-connected graph is the initial cycle of some ear decomposition~\citep{whitney}.

\begin{theorem}\label{thm:sparsedhg}
Let $G$ be a $2$-connected distance-hereditary graph with $n\geq 3$. Then $G$ is sparse if and only if $G$ is one of the following graphs: triangle, square, diamond, $K_{2,3}$, $K_{2,4}$, or $K_{3,3}^-$.
\end{theorem}

{\bf Proof.} \ If $G$ is a triangle, square, diamond, $K_{2,3}$, $K_{2,4}$, or $K_{3,3}^-$ then $G$ is sparse. Conversely, suppose that $G$ is sparse, and let $G_0\cup G_1\ldots\cup G_p$ be an ear decomposition of $G$. By Lemma~\ref{lem:possible-cycles}, $G_0$ is a triangle, a square, or a $6$-cycle. We analyze below the possible cases for $G_0$.

Henceforth, whenever the arguments used in the proof lead to the existence of a bad subgraph (contradicting the sparseness of $G$) or a $k$-cycle for $k=5$ or $k\geq 7$ (contradicting Lemma~\ref{lem:possible-cycles}), we will simply use an (*) to indicate the contradiction, in order to shorten the explanation.

$\blacktriangleright$ \ $G_0$ is a triangle: If $p=0$ then $G$ is a triangle. If $p\geq 1$, we analyze the possible cases for $G_1$.

If $G_1=P_3$ then $G_0\cup G_1$ is a diamond. If $G_1=P_4$ or $G_1=P_5$ then $G$ contains a $5$-cycle (*). Finally, if $G_1=P_k$ for $k\geq 6$ then $G$ contains a $(k+1)$-cycle (*).

If $p\geq 2$, recall that $G_0\cup G_1$ is a diamond. If $G_2=P_2$ then $G$ contains a \mbox{$K_4$ (*)}. If $G_2=P_3$ then $G_0\cup G_1\cup G_2$ is bad (*). If $G_2=P_4$ then $G$ contains a $5$-cycle, no matter the endpoints of $G_2$ are adjacent or not (*). If $G_2=P_5$ and the endpoints of $G_2$ are adjacent then $G$ contains a $5$-cycle (*). If $G_2=P_5$ and the endpoints of $G_2$ are not adjacent then $G$ contains a \mbox{$6$-cycle} requiring two chords, implying the existence of a bad subgraph in $G$ (*). Finally, if $G_2=P_k$ for $k\geq 6$ then $G$ contains a $(k+1)$-cycle (*).

This concludes the first case: if $G_0$ is a triangle then $G$ is either a triangle or a diamond.

$\blacktriangleright$ \ $G_0$ is a square: If $p=0$ then $G$ is a square. If $p\geq 1$, we analyze the possible cases for $G_1$.

If $G_1=P_2$ then $G_0\cup G_1$ is a diamond and, from the argumentation of the previous case, no additional ears can exist, i.e., $p=1$ and $G$ is a diamond. If $G_1=P_3$ and the endpoints of $G_1$ are adjacent then $G$ contains a $5$-cycle (*). If $G_1=P_3$ and the endpoints of $G_1$ are not adjacent then $G_0\cup G_1$ is a $K_{2,3}$. If $G_1=P_4$ and the endpoints of $G_1$ are adjacent then $G_0\cup G_1$ is a domino, that requires an additional edge to form a $K_{3,3}^-$; thus there must be an additional ear, and the analysis is postponed. If $G_1=P_4$ and the endpoints of $G_1$ are not adjacent then $G$ contains a $5$-cycle (*). If $G_1=P_5$ and the endpoints of $G_1$ are adjacent then $G$ contains a $5$-cycle and a $7$-cycle (*). If $G_1=P_5$ and the endpoints of $G_1$ are not adjacent then $G$ contains a $6$-cycle requiring two chords, implying the existence of a bad subgraph in $G$ (*). Finally, if $G_1=P_k$ for $k\geq 6$ then $G$ contains a $(k+1)$-cycle or $(k+2)$-cycle, depending on the adjacency relation between the endpoints of $G_1$ (*).

Thus, if $G_0$ is a square and $p\geq 1$ then $G_0\cup G_1$ is either a $K_{2,3}$ or a domino. These subcases are analyzed below.

$\blacktriangleright\blacktriangleright$ \ $G_0\cup G_1$ is a $K_{2,3}$: If $p=1$ then $G$ is a $K_{2,3}$. If $p\geq 2$, we analyze the possible cases for $G_2$. Assume that $V(G_0\cup G_1)$ is partitioned into stable sets $\{u,v\}$ and $\{x,y,z\}$.

Note that $G_2$ cannot be a $P_2$, otherwise $G_0\cup G_1\cup G_2$ is bad (*). If $G_2=P_3$ and the endpoints of $G_2$ are adjacent then $G$ contains a $5$-cycle (*). If $G_2=P_3$ and the endpoints of $G_2$ are $u$ and $v$ then $G$ contains a $K_{2,4}$. If $G_2=P_3$ and the endpoints of $G_2$ are in $\{x,y,z\}$ then $G_0\cup G_1\cup G_2$ is a $K_{3,3}^-$. If $G_2=P_4$ and the endpoints of $G_2$ are adjacent then $G_0\cup G_1\cup G_2$ contains a $6$-cycle requiring an additional chord; but this would imply the existence of seven vertices and ten edges in $G_0\cup G_1\cup G_2$ (*). If $G_2=P_4$ and the endpoints of $G_2$ are not adjacent then $G$ contains a $5$-cycle. If $G_2=P_5$ and the endpoints of $G_2$ are $u$ and $v$ then $G_0\cup G_1\cup G_2$ contains a $6$-cycle which requires two additional chords; thus $G_0\cup G_1\cup G_2$ contains eight vertices and at least twelve edges (*). If $G_2=P_5$ and the endpoints of $G_2$ are not both in $\{u,v\}$ then a $k$-cycle is formed for \mbox{$k\geq 7$ (*)}. Finally, if $G_2=P_k$ for $k\geq 6$ then a $(k+1)$-cycle is formed for \mbox{$k\geq 6$ (*)} (this conclusion holds for any possible pair of endpoints of $G_2$ in $V(G_0\cup G_1)$).

Thus, if $G_0\cup G_1$ is a $K_{2,3}$ and $p\geq 2$ then $G_0\cup G_1\cup G_2$ is either a $K_{2,4}$ or a $K_{3,3}^-$. If $p=2$ then $G$ is either the graph $K_{2,4}$ or the graph $K_{3,3}^-$. If $p\geq 3$, we analyze the possibilities for $G_3$.

We first observe that if $G_0\cup G_1\cup G_2$ is a $K_{2,4}$ and the new ear $G_3$ is a $P_3$ then $G_0\cup G_1\cup G_2\cup G_3$ is bad (*). In all the remaining possibilities, we obtain the same contradictions as those previously obtained when ear $G_2$ was added to $G_0\cup G_1$, because we can always look at the endpoints of $G_3$ as if they were located in a $K_{2,3}$. Therefore, assume that $G_0\cup G_1\cup G_2$ is a $K_{3,3}^-$.

Note that $G_3$ cannot be a $P_2$, otherwise $G_0\cup G_1\cup G_2\cup G_3$ is bad (*).  If $G_3=P_3$ then $G_0\cup G_1\cup G_2\cup G_3$ has seven vertices and ten edges, i.e., it is bad (*). If $G_3=P_4$, let $G_0\cup G_1\cup G_2$ consisting of a cycle $v_1v_2v_3v_4v_5v_6v_1$ together with the $3$-chords $v_1v_4$ and $v_2v_5$. Also, let $G_3=v_iu_1u_2v_j$, where $v_i$ and $v_j$ are the endpoints of $G_3$ and $i<j$.

Assume first that $v_i=v_1$ (by symmetry, this case is the same for any choice of $v_i$ in $\{v_1,v_2,v_4,v_5\}$). If the endpoints of $G_3$ are $v_1$ and $v_2$ then $G$ contains the $8$-cycle $v_1u_1u_2v_2v_3v_4v_5v_6v_1$ (*). If the endpoints of $G_3$ are $v_1$ and $v_3$ then $G$ contains the $5$-cycle $v_1u_1u_2v_3v_2v_1$ and the $7$-cycle $v_1u_1u_2v_3v_4v_5v_6v_1$ (*). If the endpoints of $G_3$ are $v_1$ and $v_4$ then $G$ contains the $6$-cycle $v_1v_2v_3v_4u_2u_1v_1$, that requires two chords (note that there is already one chord, namely $v_1v_4$); but this would imply the existence of eight vertices and thirteen edges in $G_0\cup G_1\cup G_2\cup G_3$ (*). If the endpoints of $G_3$ are $v_1$ and $v_5$ then $G$ contains the $7$-cycle $v_1v_2v_3v_4v_5u_2u_1v_1$ (*). If the endpoints of $G_3$ are $v_1$ an $v_6$ then $G$ contains the $8$-cycle $v_1v_2v_3v_4v_5v_6u_2u_1v_1$ (*).

Next, assume that $v_i\in\{v_3,v_6\}$. In fact, the only case that remains to be analyzed is $G_3=v_3u_1u_2v_6$. But then a $6$-cycle $v_3u_1u_2v_6v_1v_2v_3$ is formed that needs two additional chords, leading to a contradiction -- the existence of eight vertices and thirteen edges in $G_0\cup G_1\cup G_2\cup G_3$ (*).

Finally, if $G_3=P_k$ for $k\geq 5$, whatever are the endpoints of $G_3$ a $k$-cycle is formed for $k=5$ or \mbox{$k\geq 7$ (*)}.

Thus, if $G_0\cup G_1\cup G_2$ is a $K_{3,3}^-$ then $G$ is the graph $K_{3,3}^-$. To conclude this subcase, if $G_0\cup G_1$ is a $K_{2,3}$ then $G$ is a $K_{2,3}$, a $K_{2,4}$, or a $K_{3,3}^-$.

$\blacktriangleright\blacktriangleright$ \ $G_0\cup G_1$ is a domino: In this case $G_0\cup G_1$ requires an additional edge between two of its vertices to form a $K_{3,3}^-$. Thus, we may assume without loss of generality that $G_2=P_2$ and $G_0\cup G_1\cup G_2$ is a $K_{3,3}^-$. But, from the argumentation of the previous subcase, no additional ears can exist. Then $p=2$ and $G$ is the graph $K_{3,3}^-$, and this completes the proof of this subcase.

$\blacktriangleright$ \ $G_0$ is a $6$-cycle: Note that two chords must be added to $G_0$ in order to form a $K_{3,3}^-$. Thus, we may assume without loss of generality that $G_1=P_2$, $G_2=P_2$, and $G_0\cup G_1\cup G_2$ is a $K_{3,3}^-$. As already explained, no additional ears can exist. Then $p=2$ and $G$ is the graph $K_{3,3}^-$. This concludes the proof of the theorem. \ $\Box$

\begin{theorem}\label{thm:mdecdhg}
Let $G$ be a $2$-connected distance-hereditary graph. Then the following are equivalent:
\begin{itemize}
\setlength\itemsep{0em}
\item[$(a)$] $G$ is m-decyclable;
\item[$(b)$] $G$ is sparse and $K_{2,4}$-free;
\item[$(c)$] $G$ is one of the following graphs: triangle, square, diamond, $K_{2,3}$, \mbox{or $K_{3,3}^-$.}
\end{itemize}
\end{theorem}

{\bf Proof.}

$(a)\Rightarrow (b)$. Follows from Propositions~\ref{prop:super-prop}(g) and~\ref{prop:super-prop}(h).

$(b)\Rightarrow (c)$. Follows from Theorem~\ref{thm:sparsedhg}.

$(c)\Rightarrow (a)$. All the graphs listed in item $(c)$ are m-decyclable. \ $\Box$

\begin{corollary}
Let $G$ be a $2$-connected, $K_{2,4}$-free distance-hereditary graph. Then $G$ is matching-decyclable if and only if $G$ is sparse.
\end{corollary}

If $G$ is a (not necessarily $2$-connected) m-decyclable distance-hereditary graph then, by Theorem~\ref{thm:mdecdhg}, each block of $G$ is a bridge, triangle, square, diamond, $K_{2,3}$, or $K_{3,3}^-$. However, as in the case of chordal graphs, some subgraph configurations are forbidden for $G$. They can be determined by combining the possible blocks into minimal subgraphs that are not m-decyclable. Instead of a exhaustive description of such forbidden configurations, we present an $O(n)$-time algorithm for the recognition of $m$-decyclable distance-hereditary graphs.

\begin{theorem}\label{thm:recog-mdecdhg}
M-decyclable distance-hereditary graphs can be recognized in $O(n)$ time.
\end{theorem}

{\bf Proof.} \ The proof is similar to the proof of Theorem~\ref{thm:recog-mdecchordal}. Given a distance-herditary graph $G$, check whether $m\leq\lfloor\,\frac{3}{2}n\,\rfloor-1$. If so, $m=O(n)$ and then the block decomposition of $G$ can be obtained in $O(n)$ time. Next, check whether every block of the decomposition is a bridge, triangle, square, diamond, $K_{2,3}$, or $K_{3,3}^-$; this can be easily done in $O(1)$ time per block since the number of vertices in a block is at most six. Then, execute the following steps: (1) remove the bridges; (2) set all the remaining edges as ``unmarked''; (3) take any leaf block $B$ and try to find a decycling set $M_B$ of $B$ formed only by unmarked edges \ -- \ if not possible, stop ($G$ is not m-decyclable); (4) adjust $M_B$ so that its edges are not incident to the cut vertex $x_B$ of $B$ (if possible); (5) if $x_B$ is covered by $M_B$, mark all the edges incident to $x_B$ in $G-B$; (6) remove $V(B)\setminus\{x_B\}$ from the graph; (7) if no more edges are left, stop and return $\bigcup \ \{M_B: B \ \mbox{is a block}\}$ as a decycling set of $G$, otherwise go back to step (3).  The entire process clearly runs in $O(n)$ time. \ $\Box$

\medskip

An {\em md-star} is a graph $G$ such that: (a) $G$ contains exactly one cut vertex $x$; (b) $G$ contains at most one diamond block, and the remaining blocks are bridges or triangles; (c) if there is a diamond block $D$ then $d_D(x)=3$.

Let $G$ be an m-decyclable, nontrivial connected cograph. Since every cograph is distance-hereditary, by Theorem~\ref{thm:mdecdhg} and the fact that $K_{3,3}^-$ is not a cograph, it follows that every block of $G$ is a bridge, triangle, square, diamond, or $K_{2,3}$. In addition, since $G$ is $P_4$-free, $G$ contains at most one cut-vertex. If $G$ contains no cut-vertex then $G$ is a $K_2$, triangle, square, diamond, or $K_{2,3}$. If $G$ contains exactly one cut vertex $x$ then no block of $G$ is a square, a $K_{2,3}$, or a diamond $D$ with $d_D(x)=2$, otherwise there would be an induced $P_4$ in $G$; in addition, $G$ contains at most one diamond, otherwise $G$ would have a bad subgraph; thus, $G$ is an md-star. Hence:

\begin{corollary}
Let $G$ be a nontrivial connected cograph. Then $G$ is m-decycla\-ble if and only if $G$ is one of the following graphs: $K_2$, triangle, square, diamond, $K_{2,3}$, md-star.
\end{corollary}

\section{Conclusions}

In this work we considered the following question: characterize matching-decyclable graphs belonging to a special class $\C$. This question was solved for chordal graphs, split graphs, distance-hereditary graphs, and  cographs. In such classes (except in distance-hereditary graphs) being matching-decyclable is equivalent to being sparse. In addition, the presented characterizations lead to simple $O(n)$-time recognition algorithms.

The graph $K_{2,4}$ is sparse but not m-decyclable. Hence, in planar graphs, being m-decyclable is not equivalent to being sparse. An interesting question is to find a subclass of planar graphs in which these concepts are equivalent.

Finally, for Hamiltonian subcubic graphs, we proved that deciding whether a Hamiltonian fairly cubic graph is m-decyclable is NP-complete. This leads to an interesting by-product: deciding whether a Hamiltonian fairly cubic graph contains a Hamiltonian path whose endpoints are the vertices of degree two is NP-complete.

\if 10

\section*{Appendix}

In this appendix we briefly explain how the computer program (written in C) checks the validation of Properties 1 to 6 stated in the proof of Theorem~\ref{thm:npc-cubic}. The program takes as input: (a) a graph $G$ with $n$ vertices ($n\leq 20$), numbered $0,1,2,\ldots,n-1$, and $m$ edges; (b) a list of terminals (as explained in the proof of Theorem~\ref{thm:npc-cubic}). Then it first lists all the Hamiltonian paths (if any) between pairs of terminals, and next it checks Property 6 by listing all the partitions $(X,Y)$ of $V(G)$ and testing whether both $X$ and $Y$ form paths between terminals (of course, the test is made only if both $X$ and $Y$ contain at least two terminals each).

Our focus is the gadget $G_i$ in Figure~\ref{fig:gadget-red2}(a). Suppose we number the vertices of $G_i$ as shown in Figure~\ref{fig:gadget-numbered} (diamonds can be viewed as vertices of degree two, as explained in the proof of Theorem~\ref{thm:npc-cubic}). Following this numbering, we show below an execution of the program.

\begin{figure}[htbp]
\centering
\includegraphics[width=\textwidth]{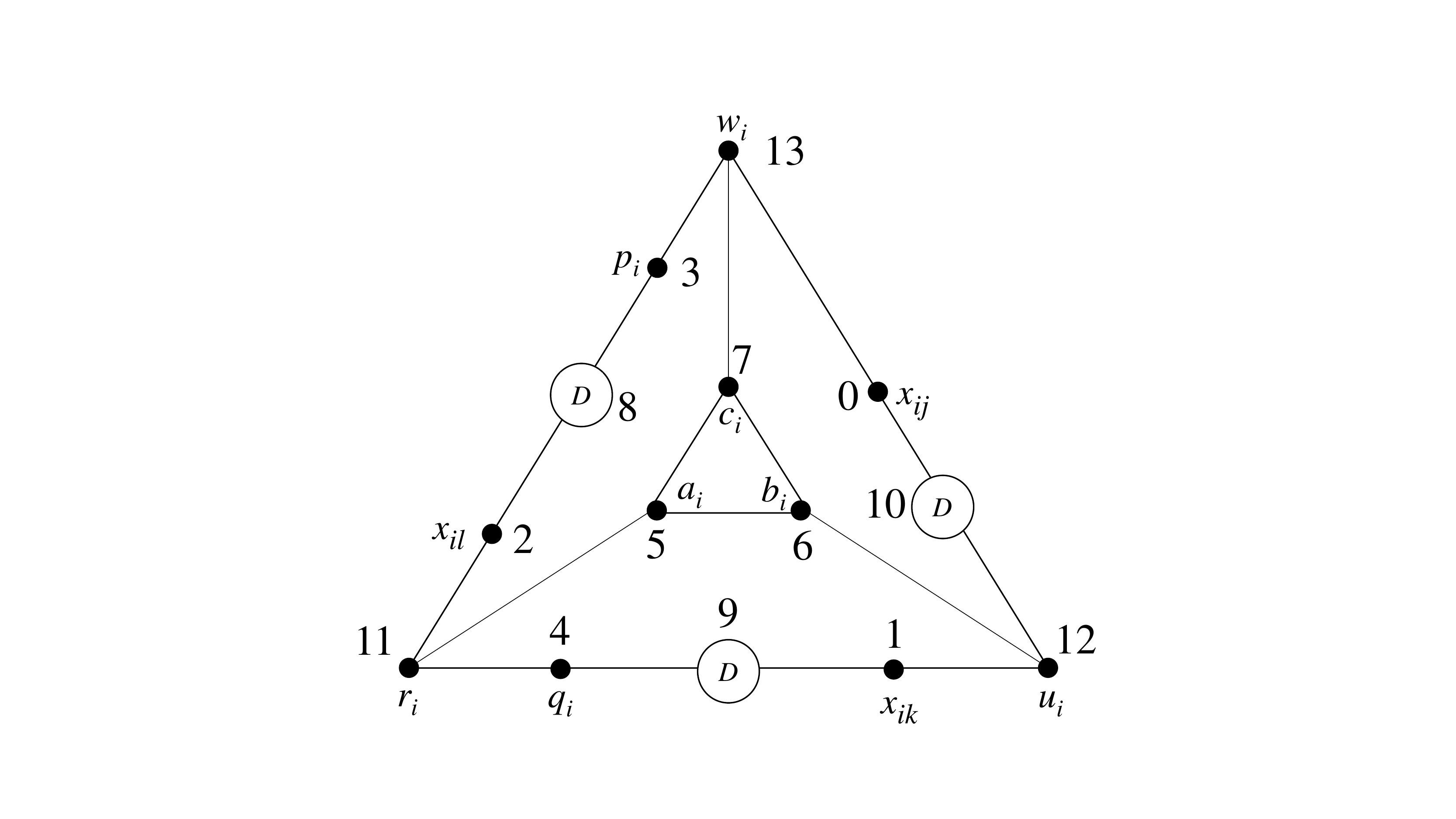}
\vspace{-1.5cm}
\caption{Numbering the vertices of gadget $G_i$.}\label{fig:gadget-numbered}
\end{figure}

\begin{tt}
------------------------------------------------\\
Enter the value of n (max n=20): 14\\
Enter the value of m: 17\\
------------------------------------------------\\
Enter the edges:\\
Edge: 0 10\\
Edge: 0 13\\
Edge: 1 9\\
Edge: 1 12\\
$\vdots$\\
------------------------------------------------\\
Enter the number of terminals: 5\\
Enter the list of terminals: 0 1 2 3 4\\
\hspace*{1cm}\\
Hamiltonian Path: 0 10 12 1 9 4 11 5 6 7 13 3 8 2\\
Hamiltonian Path: 0 10 12 6 5 7 13 3 8 2 11 4 9 1\\
Hamiltonian Path: 1 9 4 11 2 8 3 13 7 5 6 12 10 0\\
Hamiltonian Path: 1 9 4 11 5 7 6 12 10 0 13 3 8 2\\
Hamiltonian Path: 2 8 3 13 0 10 12 6 7 5 11 4 9 1\\
Hamiltonian Path: 2 8 3 13 7 6 5 11 4 9 1 12 10 0\\
Hamiltonian Path: 3 8 2 11 5 6 7 13 0 10 12 1 9 4\\
Hamiltonian Path: 4 9 1 12 10 0 13 7 6 5 11 2 8 3\\

Press enter to continue...
\end{tt}

\medskip

For each terminal vertex $j$ the program lists all the Hamiltonian paths starting at $j$ and ending at another terminal. Thus, each Hamiltonian path is printed twice. Note that the type-$x$ terminals of $G_i$ are numbered $0$, $1$, and $2$, while the other terminals are numbered 3 and 4. The program finds a unique Hamiltonian path between $0$ and $1$, and it is precisely the path $Q_i$. Thus, Property 1 is satisfied, and it is easy to see from above that Properties 2, 3, and 4 are also satisfied. Moreover, no additional Hamiltonian paths are found by the program, i.e., Property 5 is satisfied.

The program continues by listing all the partitions $(X,Y)$ of $V(G_i)$ such that both $X$ and $Y$ contain at least two terminals each, and tests whether both $X$ and $Y$ form paths. An example is given below.

\begin{tt}
------------------------------------------------\\
Partitions: \\
$\vdots$\\
------------------------------------------------\\
1st subset: 7 6 5 4 3 0 10 11 12 13\\
2nd subset: 9 8 2 1\\
Hamiltonian Path: 4 11 5 7 6 12 10 0 13 3\\
------------------------------------------------\\
$\vdots$\\
\end{tt}

\vspace{-0.4cm}
The above partition is listed because the first subset contains the terminals $0$, $3$, and $4$, and the second contains the terminals $1$ and $2$. In this example, the first subset induces a subgraph with a Hamiltonian path between the terminals $3$ and $4$, but the second induces a subgraph with no path between terminals. Hence, this partition is not a counter-example to Property 6. The program is designed to stop the execution as soon as it finds a partition $(X,Y)$ such that both $X$ and $Y$ form paths between terminals. However, it lists all the possible partitions and no counter-example is found. Hence, Property 6 is satisfied.

The program (font and executable files) can be found at

\begin{center}
{\tt www.ic.uff.br\textbackslash\textasciitilde fabiop\textbackslash hamiltonian\textunderscore test.c}\\
{\tt www.ic.uff.br\textbackslash\textasciitilde fabiop\textbackslash hamiltonian\textunderscore test.exe}
\end{center}

The output of the program (for the above input) can be found at

\begin{center}
{\tt www.ic.uff.br\textbackslash\textasciitilde fabiop\textbackslash output.txt}
\end{center}

The executable file runs on Windows platforms, but since we employed a standard C code to develop the program, it can be easily compiled to run on other settings.

\fi

\acknowledgements
\label{sec:ack}
The authors are partially supported by CNPq and FAPERJ, Brazilian research agencies.

\nocite{*}
\bibliographystyle{abbrvnat}
\bibliography{bib-dmtcs}

\begin{thebibliography}{11}
\providecommand{\natexlab}[1]{#1}
\providecommand{\url}[1]{\texttt{#1}}
\expandafter\ifx\csname urlstyle\endcsname\relax
  \providecommand{\doi}[1]{doi: #1}\else
  \providecommand{\doi}{doi: \begingroup \urlstyle{rm}\Url}\fi

\bibitem[Bandelt and Mulder(1986)]{bandelt86}
H.-J. Bandelt and H.~M. Mulder.
\newblock Distance-hereditary graphs.
\newblock \emph{Journal of Combininatorial Theory Ser. B}, 41:\penalty0
  182--208, 1986.

\bibitem[Berge(1957)]{berge57}
C.~Berge.
\newblock Two theorems in graph theory.
\newblock \emph{Proceedings of the National Academy of Sciences of the United
  States of America}, 43\penalty0 (9):\penalty0 842--844, 1957.

\bibitem[Chae et~al.(2007)Chae, Palmer, and Robinson]{chaea-et-al-07}
G.-B. Chae, E.~M. Palmer, and R.~W. Robinson.
\newblock Counting labeled general cubic graphs.
\newblock \emph{Discrete Applied Mathematics}, 307:\penalty0 2979--2992, 2007.

\bibitem[Corneil et~al.(1981)Corneil, Lerchs, and Burlingham]{corneil}
D.~G. Corneil, H.~Lerchs, and L.~S. Burlingham.
\newblock Counting labeled general cubic graphs.
\newblock \emph{Discrete Applied Mathematics}, 3:\penalty0 163--174, 1981.

\bibitem[F{\"o}ldes and Hammer(1977)]{fh77}
S.~F{\"o}ldes and P.~L. Hammer.
\newblock Split graphs.
\newblock In \emph{Proceedings of the Eighth Southeastern Conference on
  Combinatorics, Graph Theory and Computing}, volume~19 of \emph{Congressus
  Numerantium}, pages 311--315, Louisiana State Univ., Baton Rouge, La., 1977.
  Winnipeg: Utilitas Math.

\bibitem[Garey et~al.(1976)Garey, Johnson, and Tarjan]{gjt76}
M.~R. Garey, D.~S. Johnson, and R.~E. Tarjan.
\newblock The planar hamiltonian circuit problem is np-complete.
\newblock \emph{SIAM Journal on Computing}, 5\penalty0 (4):\penalty0 704--714,
  1976.

\bibitem[Groshaus et~al.(2011)Groshaus, Hell, Klein, Nogueira, and
  Protti]{groshaus11}
M.~Groshaus, P.~Hell, S.~Klein, L.~T. Nogueira, and F.~Protti.
\newblock Cycle transversals in bounded degree graphs.
\newblock \emph{Discrete Mathematics and Theoretical Computer Science},
  13\penalty0 (1):\penalty0 45--66, 2011.

\bibitem[Howorka(1981)]{howorka81}
E.~Howorka.
\newblock A characterization of ptolemaic graphs.
\newblock \emph{J. Graph Theory}, 5:\penalty0 323--331, 1981.

\bibitem[Karp(1972)]{karp72}
R.~M. Karp.
\newblock Reducibility among combinatorial problems.
\newblock In \emph{Proc. Sympos. IBM Thomas J. Watson Res. Center}, Complexity
  of Computer Computations, pages 85--103, Yorktown Heights, N.Y., New York,
  1972. Plenum.

\bibitem[Lima et~al.(2017)Lima, Rautenbach, Souza, and Szwarcfiter]{lima16}
C.~V. G.~C. Lima, D.~Rautenbach, U.~S. Souza, and J.~L. Szwarcfiter.
\newblock Decycling with a matching.
\newblock \emph{Information Processing Letters}, 124:\penalty0 26--29, 2017.

\bibitem[Whitney(1932)]{whitney}
H.~Whitney.
\newblock Non-separable and planar graphs.
\newblock \emph{Transactions of the American Mathematical Society},
  34:\penalty0 339--362, 1932.

\end{thebibliography}


\parskip=2pt
\label{sec:biblio}

\end{document}